\documentstyle[12pt,psfig]{article}

\newcommand{\sect}[1]{\section{#1}}

\renewcommand{\theequation}{\arabic{section}.\arabic{equation}}
\newcommand{\lbl}[1]{\label{eq:#1}}
\newcommand{ \rf}[1]{(\ref{eq:#1})}
\newcommand{\vs}[1]{\rule[- #1 mm]{0mm}{#1 mm}}
\newskip\humongous \humongous=0pt plus 1000pt minus 1000pt

\newif\ifdtup

\newcommand{\eq}{\vs{2}\begin{equation}}
\newcommand{\en}{\\[2mm]\end{equation}}
\newcommand{\bea}{\begin{eqnarray}}
\newcommand{\ena}{\end{eqnarray}}
\newcommand{\lapprox}{%
\mathrel{%
\setbox0=\hbox{$<$}
\raise0.6ex\copy0\kern-\wd0
\lower0.65ex\hbox{$\sim$}
}}
\newcommand{\gapprox}{%
\mathrel{%
\setbox0=\hbox{$>$}
\raise0.6ex\copy0\kern-\wd0
\lower0.65ex\hbox{$\sim$}
}}

\def\Fpi{F_{\pi}}

\def\mhat{{\widehat m}}

\def\Mpin{M_{\pi^0}^2}
\def\Mpic{M_{\pi^{\pm}}^2}
\def\Jpm{{\bar J}_{+-}}
\def\Joo{{\bar J}_{00}}
\def\kbar{{\bar k}}

\newcommand{\NP}[1]{Nucl.\ Phys.\ {\bf #1}}

\newcommand{\PL}[1]{Phys.\ Lett.\ {\bf #1}}

\newcommand{\AN}[1]{Ann. Phys. {\bf #1}}

\newcommand{\PRev}[1]{Phys.\ Rev.\ {\bf #1}}
\newcommand{\PRL}[1]{Phys.\ Rev.\ Lett.\ {\bf #1}}

\begin{document}

\font\fifteen=cmbx10 at 15pt
\font\twelve=cmbx10 at 12pt

\begin{titlepage}

\begin{center}

\renewcommand{\thefootnote}{\fnsymbol{footnote}}

{\twelve Centre de Physique Th\'eorique\footnote{
Unit\'e Propre de Recherche 7061
}, CNRS Luminy, Case 907}

{\twelve F-13288 Marseille -- Cedex 9}

\vspace{1 cm}

{\fifteen Virtual Photons in Low Energy {\Large{\mbox{$\pi-\pi$}}} Scattering \\
}

\vspace{0.3 cm}

\setcounter{footnote}{0}
\renewcommand{\thefootnote}{\arabic{footnote}}

{\bf 
Marc KNECHT\footnote{
E-mail address~:~knecht@cpt.univ-mrs.fr} 
and Res URECH\footnote{
Institut f{\"u}r Theoretische Teilchenphysik, Universit{\"a}t Karlsruhe, D-76128 Karlsruhe.

$\ \ $Now at SBC, Hochstrasse 16, CH-4002 Basel.} 
}

\vspace{2,3 cm}

{\bf Abstract}

\end{center}

The low energy scattering of pions is investigated in the presence of electromagnetic interactions at leading order and at next-to-leading order for the amplitudes involving at most one pair of charged pions. The size of the electromagnetic corrections to the S-wave scattering lengths is found to be comparable to the size of the two loop strong interaction corrections.

\vspace{6 cm}

\noindent Key-Words: Chiral perturbation theory, Pion, Electromagnetic 
corrections.

\bigskip


\bigskip

\bigskip

\noindent August 1997

\noindent CPT-97/P.3524

\bigskip

\noindent anonymous ftp or gopher: cpt.univ-mrs.fr

\renewcommand{\thefootnote}{\fnsymbol{footnote}}

\end{titlepage}

\begin{titlepage}

$\ $

\end{titlepage}

\setcounter{footnote}{0}
\renewcommand{\thefootnote}{\arabic{footnote}}

\sect{Introduction}

Over the last few years, low energy $\pi-\pi$ scattering has been increasingly recognized as providing a perhaps  unique window through which one might contemplate the actual mechanism of spontaneous breakdown of chiral symmetry in QCD (for recent discussions, see \cite{rev}). It is nowadays widely expected that this spontaneous breakdown is triggered by the formation of a large bilinear quark condensate $<0\,\vert\,{\bar q}q\,\vert\,0>\sim -(250\ {\rm MeV})^3$, which provides the main contribution to the masses of the light pseudoscalars mesons \cite{gor}
\eq
-\,{{(m_u+m_d)<0\,\vert\,{\bar q}q\,\vert\,0>}\over{\Fpi^2M_{\pi}^2}}\ \sim\ 1\ .\lbl{ratio}
\en
However, on the basis of our present theoretical understanding of non-perturbative phenomena in asymptotically free gauge theories, the situation where the condensation of quark-antiquark pairs in the vacuum is much weaker, say 
$<0\,\vert\,{\bar q}q\,\vert\,0>\sim -(100\ {\rm MeV})^3$, cannot be excluded (see for instance \cite{kneur}). It turns out that a significant deviation of the ratio \rf{ratio} from unity influences low energy $\pi-\pi$ scattering  \cite{fss}~: The weaker the condensate, the stronger do pions interact at low energies. Unfortunately, the experimental informations available at present do not reach a precision which would allow to rule out any of the possibilities mentioned above. In this context, planed experiments such as the measurement of the lifetime of $\pi^+-\pi^-$ atoms at CERN \cite{atom}, or new high statistics $K_{\ell 4}$  experiments by the E865 and KLOE collaborations, respectively at BNL and Da$\Phi$ne \cite{hbk}, are of particular interest. 

The expected precision of these forthcoming data has triggered quite some activity at the theoretical level. In particular, the $\pi-\pi$ scattering amplitude has been obtained at next-to-next-to-leading order both in the framework of generalized chiral perturbation theory \cite{kmsf}, where the ratio \rf{ratio} is kept as a free parameter, and for the special case of standard chiral perturbation theory \cite{bcegs}, where this ratio is exactly equal to one at leading order. In the former case, the whole range allowed by the present experimental value \cite{dumb} of the $I=0$ S-wave scattering length, $a_0^0=0.26\pm0.05$, is covered as the ratio \rf{ratio} is varied between 0 and 1.
In the latter case the prediction, as taken from the recent literature \cite{bcegs,gkms}, is in the range $a_0^0=0.205-0.217$, depending mainly on  the evaluation of the $O(p^4)$ counterterms (this question is presently under further investigation \cite{leut}). These last numbers, when compared to the corresponding leading order \cite{wei} and next-to-leading order values \cite{GLl2,GL}, $a_0^0=0.16$ and $a_0^0=0.20$, respectively, strongly suggest that contributions beyond two loops are negligible (this conclusion also holds for the generalized case \cite{kmsf}). 

In order to assess the prediction of the standard case as far as low energy 
$\pi-\pi$ scattering is concerned, and thus be able to detect any deviation from the strong condensate scenario \rf{ratio} once more precise data become available, it becomes mandatory to watch out for other contributions, which might compete with the two loop strong corrections discussed above. For instance, the 
aforementioned results were obtained upon neglecting all isospin breaking effects, {\it i.e.} upon ignoring the quark mass difference $m_u - m_d$, and without taking into account electromagnetic interactions. In the present paper, we are interested in the evaluation of the corrections induced by the latter, which are expected to be the dominant ones (just as they dominate, for instance, the difference between the charged and neutral pion masses \cite{GLrep,GL}, whereas $m_u\neq m_d$ only affects both the pion masses and scattering amplitudes starting from next-to-leading order). We shall work within the framework of chiral perturbation theory \cite{wei3,GLl1,GL,GL2}, extended to include electromagnetic interactions \cite{eck,res}.
The central object of our analysis is the generating functional
\eq
e^{i{\cal Z}(v_{\mu},a_{\mu},s,p,Q_L,Q_R)}\ =
\ <0 ;{\rm out}\,\vert\,0 ;{\rm in}>
_{v_{\mu},a_{\mu},s,p,Q_L,Q_R}\ ,
\en
where the expectation value is obtained by integration over gluon, quark {\it and} photon configurations, weighted by the action corresponding to the QCD 
Lagrangian with electromagnetic interactions and in the presence of the external sources $v_{\mu},a_{\mu},s,p,Q_L,Q_R$,
\bea
{\cal L} &=& {\cal L}_{\rm QCD}^0+{\cal L}^0_{\gamma}
\,+\,{\bar q}\gamma^{\mu}[v_{\mu}(x)+\gamma_5a_{\mu}(x)]q
-{\bar q}[s(x)-i\gamma_5p(x)]q
\nonumber\\
&&
\,+A_{\mu}\,{\bar q}\gamma^{\mu}\,\big\{\,Q_L(x)\left({{1-\gamma_5}\over2}\right)
+Q_R(x)\left({{1+\gamma_5}\over2}\right)\,\big\}\,q\ .
\ena
In the above expression, ${\cal L}^0_{\rm QCD}$ stands for the QCD Lagrangian with massless light quarks, while ${\cal L}^0_{\gamma}$ stands for the Maxwell Lagrangian of the photon field $A_{\mu}$. In both cases, some gauge fixing and the corresponding Faddeev-Popov ghosts are understood. The last term provides the interaction of the $N_f$ light quarks $q=u,d,...$ with the electromagnetic field once the sources $Q_L(x)$ and $Q_R(x)$ are put equal to the constant charge matrix $Q$ appropriate for $N_f$ light flavours. The structure of the generating functional ${\cal Z}$ above has been obtained to one loop order in \cite{res} for $N_f=3$ within a systematic framework which combines the usual chiral expansion \cite{wei3,GLl1,GL,GL2} with the expansion in powers of the electromagnetic coupling $e$ (see also \cite{neu}). For our purpose, we need only to consider the case $N_f=2$. However, for those aspects which are independent of the actual number of light flavours involved, we shall quote the results for arbitrary $N_f$. Electromagnetic corrections in the two flavour case within this systematic framework have also been considered recently in \cite{ulf,ulf2}, and we shall compare with our present work in due course. Let us mention, however, that the only application that was discussed in \cite{ulf,ulf2} is the scattering of neutral pions, where photon exchanges between pions do not contribute before the two loop order. In the meantime, the formalism of \cite{ulf,ulf2} has been used to compute the corrections to the lowest order formula \cite{atom2} for the lifetime of pionium in Ref. \cite{js} (for other recent approaches to this problem, see \cite{atom3}). This computation obviously involves the scattering amplitude $\pi^+\pi^-\rightarrow\pi^0\pi^0$, but, for the purposes of the bound state problem, at an off-shell point, whereas we shall consider the same amplitude for free, on-shell, pions. Whenever comparison was possible, we found agreement between our expressions and the results of \cite{js}. Finally, electromagnetic corrections to low energy $\pi-\pi$ scattering within different contexts and/or frameworks have been considered previously in Refs. \cite{rs}, \cite{wic} and \cite{mw}. Unless otherwise stated, we shall work within the framework of standard chiral perturbation theory, where \rf{ratio} is assumed to hold.

The rest of the paper is organized as follows. In Section 2 we investigate lowest order electromagnetic effects on the $\pi-\pi$ scattering amplitudes and discuss the S-wave scattering lengths. Section 3 is devoted to the construction of the generating functional to one loop order. We give a complete list of counterterms for $N_f=2$ and for constant sources $Q_L(x) = Q_R(x) = Q$, and compute the corresponding divergent part of the generating functional. The amplitudes for $\pi^0\pi^0\rightarrow\pi^0\pi^0$ and $\pi^+\pi^-\rightarrow\pi^0\pi^0$ are worked out to one loop order in Section 4. Scattering lengths are discussed in Section 5. In particular, the definition of the scattering length for $\pi^+\pi^-\rightarrow\pi^0\pi^0$ requires a proper treatment of the infrared singularities which affect the amplitude. A final Section summarizes our results, and details on various technical aspects have been gathered in an Appendix.

\indent

\sect{Lowest Order Electromagnetic Corrections}
\setcounter{equation}{0}

It has been shown in \cite{res} that in presence of electromagnetism, 
a consistent expansion scheme is obtained if the electric charge $e$ is 
considered as a quantity of order $p$ in the chiral counting,
\eq
e,\,Q_L,\,Q_R\ \sim\ O(p)\ ,
\en
where $p$ denotes a typical momentum, much smaller than the scale $\Lambda_H\sim 1$ GeV at which the (non-Goldstone) hadronic bound states appear.
 At leading order 
in this combined expansion, and for an arbitrary number $N_f$ of light 
quark flavours, the generating 
functional is given by the tree graphs constructed from the effective 
lagrangian \cite{eck}
\bea
{\cal L}^{(2)} &=& {{F^2}\over 4}\,
\langle\,d^{\mu}U^+d_{\mu}U\,+\,
\chi^+U+U^+\chi\,\rangle
\nonumber\\[2mm]
&&\,-{1\over 4}F^{\mu\nu}F_{\mu\nu}\,
-\,{1\over{2a}}\,({\partial}\cdot A)^2\,
+\,C\,\langle\,Q_{R}UQ_{L}U^+\,\rangle\ .
\lbl{p2}
\ena
For the notation, we follow Ref. \cite{res}, where the transformation 
properties of the various quantities which enter \rf{p2} 
 under the chiral SU($N_f$)$\times$SU($N_f$) symmetry group
are displayed.  Besides the scalar and the pseudoscalar sources, the field $\chi$ contains also the mass term for the pseudoscalar mesons at leading order, $\chi = 2B\,{\cal M}+\cdots$, where ${\cal M}$ is the (diagonal) quark mass 
matrix. The low energy constant $B$ describes the bilinear light quark condensates,
\eq
B\ =\ -\,{{<0\,\vert\,{\bar u}u\,\vert\,0>}\over{F^2}}\ =
\ -\,{{<0\,\vert\,{\bar d}d\,\vert\,0>}\over{F^2}}\ =\ \cdots\lbl{cond}
\en
whereas $F$ is given by a two point correlation function of the vector and axial currents at vanishing momentum transfer,
\eq
F\,\delta^{ab}\ =\ {1\over{3i}}\,\int\,d^4x
<0\,\vert\,T\big\{({\bar q}_{\rm L}\gamma^{\mu}\lambda^aq_{\rm L})(x)
({\bar q}_{\rm R}\gamma_{\mu}\lambda^bq_{\rm R})(0)\big\}\,\vert\,0>\ .
\lbl{Fcorr}
\en
In both cases, $\vert\,0>$ stands for the vacuum in the chiral limit and in the absence of electromagnetism, $\vert\,0>=\vert\,\Omega >\vert_{{\cal M}=0,e=0}$. The matrices $\lambda^a/2$ are the generators of 
the corresponding SU($N_f$) flavour group.
The covariant derivative $d_{\mu}$, defined as 
\eq
d_{\mu}U\ =\ \partial_{\mu}U-i(v_{\mu}+Q_{R}A_{\mu}+a_{\mu})U
+iU(v_{\mu}+Q_{L}A_{\mu}-a_{\mu})\ ,
\en
and
the last term  of \rf{p2} contain the ``spurions'' $Q^a_{L}(x)$ and 
$Q^a_{R}(x)$, 
which play the role of sources for insertions, into the QCD Green's functions,
 of the electromagnetic vertex operators 
$A_{\mu}{\bar q}_{\rm L}\,(\lambda^a/2)\gamma^{\mu}\,q_{\rm L}$ and 
$A_{\mu}{\bar q}_{\rm R}\,(\lambda^a/2)\gamma^{\mu}\,q_{\rm R}$, 
respectively. The low energy constant $C$ gives an electromagnetic contribution to the charged pseudoscalar masses, for instance, 
\bea
&\Mpin &= (m_u+m_d)B\ ,
\nonumber\\
&\Mpic &= (m_u+m_d)B\,+\,2C\cdot{{e^2}\over{F^2}}\ ,\lbl{leadmass}
\ena  
which yields
\eq
Z\ \equiv\ {C\over{F^4}}\ ={{\Mpic-\Mpin}\over{2e^2F^2}}\ .\lbl{leadZ}
\en
For $F=92.4$ MeV\footnote{Since in this Section we work at  lowest order, we identify $F$ with the pion decay constant $\Fpi=92.4$ MeV \cite{pdg}.}, this gives $Z$=0.8. Alternatively, $C$ is given by the same correlator as in \rf{Fcorr}, convoluted with the free photon propagator function,
\eq
C\,\delta^{ab}\ =\ {1\over 2}\,\int\,d^4x D^{\mu\nu}(x)\,
<0\,\vert\,T\big\{({\bar q}_{\rm L}\gamma_{\mu}\lambda^aq_{\rm L})(x)
({\bar q}_{\rm R}\gamma_{\nu}\lambda^bq_{\rm R})(0)\big\}\,\vert\,0>\ .
\lbl{Ccorr}
\en
Finally, the penultimate 
term of \rf{p2} acts as a gauge fixing. Notice that $C$ is independent of the gauge parameter $a$, since the correlator in \rf{Ccorr}, involving conserved currents, is transverse.
Since this correlator itself is an order parameter of spontaneous breakdown of chiral symmetry, it has a smooth behaviour at short distances. This in turn allows to convert the representations \rf{Fcorr} and \rf{Ccorr} into  sum rules \cite{wei2,das} {\it via} unsubtracted dispersion relations. Upon saturating these sum rules with resonances, this yields  an independent evaluation of $C$ which is compatible with the number given above \cite{das,eck,bachir}. For our purposes, the leading order determination \rf{leadZ} will be sufficient.

In fact, for the two flavour case, the splitting of the pion masses is the only direct effect induced by the presence of $C$. This is best seen in the so called $\sigma$-model parametrization of the field $U$,
\eq
U\ =\ \sqrt{1-\langle\,\phi^2\,\rangle/2F^2}\,+\,i{{\phi}\over{F}}
\quad {\rm with}
\quad \phi\ =\ \left(
\begin{array}{cc}
\pi^0\, ,& \,-\sqrt{2}\pi^+\\
\sqrt{2}\pi^-\, ,&\,-\pi^0
\end{array}\right)\ ,\lbl{sigma}
\en 
in which the last term of \rf{p2} gives no interaction vertices for $Q_L(x)=Q_R(x)=Q$, with $Q$=$e\times$diag(${2\over 3},-{1\over 3}$) the two flavour charge matrix, 
\eq
C\,\langle\,QUQU^+\,\rangle
\ =\ 2C\cdot {{e^2}\over{F^2}}\cdot \pi^+\pi^-\ .\lbl{Cpipi}
\en
This means that 
if we introduce the ``isotriplet'' states $\vert\,\pi^a(M_a,{\vec p})>$, 
$a=1,2,3$ (we use the Condon-Shortley phase convention),
\bea
\vert\,\pi^1(M_{\pi^{\pm}},{\vec p})> &=& -{1\over\sqrt{2}}
\,\big(\,\vert\,\pi^+(M_{\pi^{\pm}},{\vec p})>\ +
\ \vert\,\pi^-(M_{\pi^{\pm}},{\vec p})>\,\big)\ ,
\nonumber\\
\vert\,\pi^2(M_{\pi^{\pm}},{\vec p})> &=& {i\over\sqrt{2}}
\,\big(\,\vert\,\pi^+(M_{\pi^{\pm}},{\vec p})>\ -
\ \vert\,\pi^-(M_{\pi^{\pm}},{\vec p})>\,\big)\ ,
\nonumber\\ 
\vert\,\pi^3(M_{\pi^0},{\vec p})> &=& \vert\,\pi^0(M_{\pi^0},{\vec p})>\ ,
\lbl{trip}
\ena
the amplitudes $A^{ab;cd}(s,t,u)$ for the processes $\pi^a\pi^b\to\pi^c\pi^d$ 
are expressed, after subtraction of the one photon exchange Born terms, by means of a single amplitude,
\eq
A^{ab;cd}(s,t,u)\ =\ \{\ A(s\vert t,u)\delta^{ab}\delta^{cd}\ +\ {\rm perm}\ \}
\ +\ O(e^2p^2,\,e^4)\ ,\lbl{iso}
\en
where, as in the absence of electromagnetism,
\eq
A(s\vert t,u)\ =\ {{s-2\mhat B}\over{F^2}}\ \lbl{amp2}
\en
at lowest order, with $\mhat = {1\over2}(m_u+m_d)$.
Several remarks are in order at this point. First, the observations contained in Eqs. \rf{Cpipi}, \rf{iso} and \rf{amp2} were made previously by several authors (see {\it e.g.} \cite{juerg} and references therein). Second, we have added a 
contribution $O(e^2p^2,e^4)$ in \rf{iso} to indicate that this representation 
might not be maintained once next-to-leading electromagnetic corrections are 
taken into account, even after suitable subtraction of the long range one photon exchange contributions. We shall come back to this point in the following 
Section. Finally, at first sight the representation \rf{iso} suggests that 
the leading electromagnetic corrections induced by the last term in \rf{p2} do 
not lead to isospin violations in $\pi-\pi$ scattering. However, and as  
explicitly indicated by the notation in \rf{trip}, the 
``isotriplet'' states have unequal masses, and these mass differences lead to isospin breaking contributions to the  amplitudes (for instance, the thresholds of various reactions can be distinct). Considering the S-wave scattering lengths, we write
\bea
&&a_0(00;00)\ =\ {\textstyle{1\over3}}(a_0^0)_{\rm str}+{\textstyle{2\over3}}(a_0^2)_{\rm str}\ +
\ \Delta a_0(00;00)
\nonumber\\
&&a_0(+0;+0)\ =\ {\textstyle{1\over2}}(a_0^2)_{\rm str}\ +
\ \Delta a_0(+0;+0)
\nonumber\\
&&a_0(+-;00)\ =\ -{\textstyle{1\over3}}(a_0^0)_{\rm str}+{\textstyle{1\over3}}
(a_0^2)_{\rm str}\ +
\ \Delta a_0(+-;00)
\nonumber\\
&&a_0(+-;+-)\ =\ {\textstyle{1\over3}}(a_0^0)_{\rm str}+{\textstyle{1\over6}}
(a_0^2)_{\rm str}\ +
\ \Delta a_0(+-;+-)
\nonumber\\
&&a_0(++;++)\ =\ (a_0^2)_{\rm str}\ +\ \Delta a_0(++;++)\ ,\lbl{adef}
\ena
where $(a_0^0)_{\rm str}$ and $(a_0^2)_{\rm str}$ denote the {\it strong} isospin $I=0$ and $I=2$ S-wave scattering lengths in the absence of electromagnetic corrections, and $\Delta a_0(ab;cd)$ are the corresponding corrections induced by electromagnetic effects. At lowest order, one has $(a_0^0)_{\rm str}=7M_{\pi}^2/32\pi F^2$ and $(a_0^2)_{\rm str}=-M_{\pi}^2/16\pi F^2$, where $M_{\pi}$ denotes the leading order pion mass for $e=0$, {\it i.e.} 
$M_{\pi}^2=2\mhat B$. From the expressions \rf{leadmass}, $M_{\pi}$ 
should therefore be 
identified with $M_{\pi^0}$ at this order. However, it has become customary 
to quote the values of the lowest order S-wave scattering lengths, obtained by Weinberg more than thirty years ago \cite{wei}, with the value of the {\it charged} pion mass $M_{\pi^{\pm}}=139.57$ MeV \cite{pdg} assigned to $M_{\pi}$, {\it i.e.}
\eq
(a_0^0)_{\rm str}\ =\ {{7\Mpic}\over{32\pi F^2}}\ =\ 0.16\ ,
\ \ (a_0^2)_{\rm str}\ =\ -{{\Mpic}\over{16\pi F^2}}\ =\ -0.045\ .\lbl{aW}
\en
Adopting this definition, we obtain the following values for the shifts in the S-wave scattering lengths ($\Delta_{\pi}\equiv \Mpic -\Mpin$)~:
\eq  
\begin{array}{ll}
\Delta a_0(00;00)\ =\ -\displaystyle{{{\Delta_{\pi}}\over{32\pi F^2}}}
\qquad &(-6.4\%) 
\\[2mm]
\Delta a_0(+0;+0)\ =\ \displaystyle{{{\Delta_{\pi}}\over{32\pi F^2}}}
\qquad &(+6.4\%)
\\[2mm]
\Delta a_0(+-;00)\ =\ -{\displaystyle{{\Delta_{\pi}}\over{32\pi F^2}}}
\qquad &(-2.1\%)
\\[2mm]
\Delta a_0(+-;+-)\ =\ {\displaystyle{{\Delta_{\pi}}\over{16\pi F^2}}}
\qquad &(+6.4\%)
\\[2mm]
\Delta a_0(++;++)\ =\ {\displaystyle{{\Delta_{\pi}}\over{16\pi F^2}}}
\qquad &(+6.4\%)\ .
\end{array}\lbl{a0}
\en
The absolute variation is at most ${{\Delta_{\pi}}/{16\pi F^2}}=0.003$. The relative variations are shown between parentheses.
Furthermore, the scattering lengths of the four amplitudes which involve at least one pair of charged pions satisfy the usual isospin relations. In order to make this apparent, we introduce the modified scattering lengths
\bea
a_0^0 &\equiv\  (a_0^0)_{\rm str}
\ +\ 5\,\displaystyle{{{\Delta_{\pi}}\over{32\pi F^2}}} &=\ 0.166
\nonumber\\[2mm]
a_0^2 &\equiv\  (a_0^2)_{\rm str}
\ +\ \displaystyle{{{\Delta_{\pi}}\over{16\pi F^2}}} &=\ -0.042\ ,
\lbl{a0corr}
\ena
and obtain
\bea
&&a_0(00;00)\ =\ {\textstyle{1\over3}}a_0^0
+{\textstyle{2\over3}}a_0^2\ -
\ \displaystyle{{{\Delta_{\pi}}\over{8\pi F^2}}}
\nonumber\\ 
&&a_0(+0;+0)\ =\ {\textstyle{1\over2}}a_0^2
\nonumber\\
&&a_0(+-;00)\ =\ -{\textstyle{1\over3}}a_0^0
+{\textstyle{1\over3}}a_0^2
\nonumber\\
&&a_0(+-;+-)\ =\ {\textstyle{1\over3}}a_0^0
+{\textstyle{1\over6}}a_0^2
\nonumber\\
&&a_0(++;++)\ =\ a_0^2\ .\lbl{acorr}
\ena
Thus, at leading order, we can distinguish two consequences of electromagnetic interactions for the $\pi-\pi$ scattering lengths~: A shift in the strong S-wave isospin scattering lengths $(a_0^I)_{\rm str}$, described by Eq. \rf{a0corr}, and an additional 
explicit isospin breaking contribution to $a_0(00;00)$ in Eq. \rf{acorr}. These corrections to the scattering lengths are comparable in magnitude to the pure strong interaction two loop effects as evaluated in the recent literature 
\cite{bcegs,gkms}.

The  expressions \rf{a0} above disagree with those obtained by Maltman and Wolfe in Ref. \cite{mw}. Of course, on general grounds \cite{haag} the result \rf{a0} does not depend on whether one uses the parametrization \rf{sigma} or the exponential  parametrization
\eq
U\ =\ e^{i\phi/F}\ ,
\en
as Maltman and Wolfe do. Rather, these authors have not taken into account the isospin violating contributions that come from the first term of \rf{p2} {\it via} the mass difference between charged and neutral pions. Therefore, their expressions for the leading order electromagnetic corrections to the scattering 
lengths are not correct.

Interestingly enough, at the same order, the relation \rf{iso} also holds in the generalized case\footnote{J.~Stern, private communication.}, and with the {\it same} expression \rf{amp2} for $A(s\vert t,u)$. The difference with the standard case comes from the fact that $2\mhat B$ no longer represents, for $e=0$, the only contribution to the pion mass even at leading order. Instead, one has
\eq
{{2\mhat B}\over{M_{\pi}^2}}\ =\ {{4-\alpha}\over3}\ ,
\en
where the parameter $\alpha$ (which is {\it not} related to $e^2/4\pi$~!) varies between 1, the standard case of a strong condensate, and 4, the extreme case where the condensate would vanish\footnote{Within the SU(3) framework, $\alpha$ can be related to the quark mass ratio $m_s/\mhat$ \cite{kmsf}~:~$\alpha =1$ corresponds to the standard value $m_s/\mhat =25.9$ at leading order \cite{GLrep,GL2}, whereas for $\alpha =4$, the quark mass ratio drops to $m_s/\mhat =6.3$ \cite{fss}.}. Correspondingly, the leading order S-wave scattering lengths become \cite{fss}, for $e=0$,
\eq
(a_0^0)_{\rm str}\ =\ {1\over{96\pi}}{{\Mpic}\over{F^2}}\,(5\alpha +16)
\ ,\ \ (a_0^2)_{\rm str}\ =\ {1\over{48\pi}}{{\Mpic}\over{F^2}}\,(\alpha -4)\ .
\en
However, the result \rf{acorr}   
for the scattering lengths holds as it stands, provided the definition of the corrected isospin scattering lengths is suitably 
modified to read
\bea
a_0^0 &\equiv\  (a_0^0)_{\rm str}
\ +\ {5\over3}\,(4-\alpha)\,\displaystyle{{{\Delta_{\pi}}\over{32\pi F^2}}}\ ,
\nonumber\\
a_0^2 &\equiv\  (a_0^2)_{\rm str}
\ +\ {2\over3}\,(4-\alpha)\,\displaystyle{{{\Delta_{\pi}}\over{32\pi F^2}}}\ .
\lbl{Gacorr}
\ena
As the condensate becomes weaker ({\it i.e.} as $\alpha$ grows towards 4), the magnitude of the correction induced by electromagnetic effects in the scattering lengths $a_0^I$ decreases, while the explicit isospin breaking component of $a_0(00;00)$ is not affected by the size of the condensate.

We conclude therefore that electromagnetic corrections to the S-wave scattering lengths are sizeable at leading order. Whereas they remain small as compared to the one loop strong interaction corrections, they are of the same  order of magnitude than the two loop corrections \cite{kmsf,bcegs,gkms}. In the sequel, we shall investigate how these numbers are affected by next-to-leading corrections. Our study will henceforth be limited to the standard case.

\indent

\sect{The Generating Functional to One Loop}
\setcounter{equation}{0}

In this Section, we construct to  one loop order the generating functional 
${\cal Z}(v_{\mu},a_{\mu},s,p,Q_L,Q_R)$ in the presence of 
electromagnetic interactions.

At next-to-leading order, this generating functional involves one loop graphs 
with vertices from ${\cal L}^{(2)}$, and tree graphs with vertices from 
${\cal L}^{(2)}$ and at most one vertex from the next-to-leading effective 
lagrangian\footnote{This decomposition of ${\cal L}^{(4)}$ corresponds to the 
minimal number of sources $Q_{L,R}$ involved. Terms proportional to the electric charge $e$ are also 
present in ${\cal L}_{p^4}$ {\it via}, for instance, the covariant derivative 
$d_{\mu}$.} 
\eq
{\cal L}^{(4)}\ =\ {\cal L}_{p^4}\ +\ {\cal L}_{e^2p^2}\ +\ {\cal L}_{e^4}\ .
\en
The first term contains the purely QCD low energy interactions among the 
pseudoscalar 
mesons. In the case of two flavours and for $e=0$, its expression was given in 
Ref. \cite{GL}. In the presence of electromagnetic interactions, it 
reads\footnote{We use here the SU(2)$\times$SU(2) formalism, rather than the 
O(4) formalism of \cite{GL}, particular to the two flavour case. We also omit 
the Wess-Zumino term, which describes anomalous couplings between photons and 
pions (for a review, see \cite{bij}).}~:
\bea
{\cal L}_{p^4} &=& {{l_1}\over 4}\,\langle\,d^{\mu}U^+d_{\mu}U\,
\rangle^2\,+\,{{l_2}\over 4}\,\langle\,d^{\mu}U^+d^{\nu}U\,
\rangle\,\langle\,d_{\mu}U^+d_{ 
\nu}U\,\rangle
\nonumber\\
&&
+\,{{l_3}\over 16}\,\langle\,\chi^+U+U^+\chi\,\rangle^2
\,+\,{{l_4}\over 4}\,\langle\,d^{\mu}U^+d_{\mu}\chi
+ d^{\mu}\chi^+d_{\mu}U\,\rangle
\nonumber\\
&&
+\,l_5\,\langle\,G^{\rm R}_{\mu\nu}UG^{{\rm L}\,\mu\nu}U^+\,\rangle
\,+\,{{il_6}\over 2}\,\langle\,G^{\rm R}_{\mu\nu}d^{\mu}Ud^{\nu}U^+ 
+ G^{\rm L}_{\mu\nu}d^{\mu}U^+d^{\nu}U\,\rangle
\nonumber\\
&&
-\,{{l_7}\over 16}\,\langle\,\chi^+U-U^+\chi\,\rangle^2
\,+\,{1\over 4}\,(h_1+h_3)\,\langle\,\chi^+\chi\,\rangle
\nonumber\\
&&
+\,{1\over 2}\,(h_1-h_3)\,Re ({\rm det}\chi )
\,-\,h_2\,\langle\,G^{\rm R}_{\mu\nu}G^{{\rm R}\,\mu\nu}
+G^{\rm L}_{\mu\nu}G^{{\rm L}\,\mu\nu}\,\rangle\ .
\ena
For the definition of the quantities $G^{R,L}_{\mu\nu}$, we refer the reader to Eqs. (A.6) and (A.13) of the Appendix.

The loop graphs with vertices from ${\cal L}^{(2)}$ generate divergences, 
which are absorbed
into the renormalization of the low energy constants of ${\cal L}^{(4)}$. In 
order to compute these divergent pieces, one may evaluate, for an arbitrary number of space-time dimensions $d$, the functional 
determinant of the quadratic part of the quantum fluctuations of the pion and 
photon fields around their classical configurations in the presence of the external sources (see Ref. \cite{res} and the Appendix).
For our purposes, we need only to consider  
the two flavour case (the relevant expressions for $N_f$ arbitrary can be found in the Appendix), and with the sources $Q_{L}(x)$ and 
$Q_{R}(x)$ put to their constant value,
\eq
Q_{L}(x)\ =\ Q_{R}(x)\ =\ Q\ \  
,\ Q\ \equiv\ e\times\,{\rm diag}\,
({2\over3},-{1\over3})\ .\lbl{const}
\en
Furthermore, unless otherwise stated, from now on we restrict ourselves to the Feynman gauge $a=1$.
Upon using the identity
\eq
\langle\,Q\,\rangle^2\ =\ {1\over5}\,\langle\,Q^2\,\rangle\ ,\lbl{trace}
\en
we then obtain, from the expressions (A.9) and (A.14) of the Appendix,
\bea
&&{\cal Z}_{\rm one\ loop,\,div}\ =
\ -\,\frac{1}{16\pi^{2}}\frac{1}{d-4}\int d^{4}x\;
\nonumber\\
&&\qquad
\bigg\{\,
{1\over 12}\,\langle\,d^{\mu} U  
^+ d_{\mu} U\,\rangle^2
+{1\over 6}\,\langle d^{\mu} U^+ d^{\nu}U\,\rangle\,
\langle\,d_{\mu} U^+ d_{\nu} U\,\rangle
\nonumber\\
&&\qquad
-{1\over{32}}\,\langle\,\chi^+U+U^+\chi\,\rangle^2
+{1\over 2}\,\langle\,d^{\mu}U^+d_{\mu}\chi 
+ d^{\mu}\chi^+d_{\mu}U\,\rangle
\nonumber\\
&&\qquad
-{1\over 6}\,\langle\,G^R_{\mu\nu} U G^{L\,\mu\nu}U^+\,\rangle
-{i\over 6}\,\langle\,G^R_{\mu\nu} d^{\mu} U d^{\nu} U^+ +
G^L_{\mu\nu} d^{\mu} U^+ d^{\nu} U\,\rangle
\nonumber\\
&&\qquad
+{1\over 2}\,\langle\,\chi^+\chi\,\rangle
-{1\over 12}\,\langle\, G^R_{\mu\nu} G^{R\,\mu\nu} +
G^L_{\mu\nu} G^{L\,\mu\nu}\,\rangle
+Re ({\rm det}\chi) 
\nonumber\\
&&\qquad
+{1\over 30}\,\langle\,Q^2\,\rangle\,F^{\mu\nu}F_{\mu\nu}
\nonumber\\
&&\qquad
-\bigg({27\over 20}+{Z\over 5}\bigg)F^2
\,\langle\,d^{\mu} U^+ d_{\mu} U\,\rangle
\,\langle\,Q^2\,\rangle
+2ZF^2\,\langle\,d^{\mu}U^+d_{\mu}U\,\rangle
\langle\,QUQU^+\,\rangle
\nonumber\\
&&\qquad
-{{3F^2}\over 4}\,
\big(\,\langle\,d^{\mu}U^+QU\,\rangle\,
\langle\,d_{\mu}U^+QU\,\rangle\,+
\,\langle\,d^{\mu}U^+QU^+\,\rangle\,
\langle\,d_{\mu}UQU^+\,\rangle\,\big)
\nonumber\\  
&&\qquad
+2ZF^2\,\langle\,d^{\mu}U^+QU\,\rangle\,
\langle\,d_{\mu}UQU^+\,\rangle
\nonumber\\
&&\qquad
-\bigg({1\over 4}+{Z\over 5}\bigg)F^2\,
\langle\,\chi^+U+U^+\chi\,\rangle\,\langle\,Q^2\,\rangle
\nonumber\\
&&\qquad
+\bigg({1\over 4}+2Z\bigg)F^2
\,\langle\,\chi^+U+U^+\chi\,\rangle\,
\langle\,QUQU^+\,\rangle
\nonumber\\
&&\qquad
+\bigg({1\over 8}-Z\bigg)F^2
\,\langle\,(\chi U^+ - U\chi^+ )QUQU^+
+(\chi^+U-U^+\chi )QU^+QU\,\rangle
\nonumber\\  
&&\qquad
+{{F^2}\over 4}\,\langle\,d_{\mu}U^+[(c_R^{\mu}Q),Q]U
+d_{\mu}U[(c_L^{\mu}Q),Q]U^+\,\rangle
\nonumber\\
&&\qquad
+\bigg({3\over 2}+3Z+12Z^2\bigg)F^4\,
\langle\,QUQU^+\,\rangle^2
\nonumber\\
&&\qquad
-\bigg(3+{{3Z}\over 5}+{{12Z^2}\over 5}\bigg)F^4
\,\langle\,QUQU^+\,\rangle\,\langle\,Q^2\,\rangle
\nonumber\\
&&\qquad
+\bigg( {3\over 2}-{{12Z}\over 5}+{{84Z^2}\over 25}\bigg)F^4
\,\langle\,Q^2\,\rangle^2\ \bigg\}\ .
\lbl{div}
\ena 
The covariant derivatives $(c^{\mu}_{R,L}Q)$ are defined in the Appendix, Eq. (A.12).

The complete list of possible counterterms in ${\cal L}_{e^2p^2}$ and 
${\cal L}_{e^4}$ was given for $N_f=3$ in Refs. \cite{res,neu}. For $N_f=2$, the number of possibilities decreases, due to additional trace identities for products of 2$\times$2 matrices. 
In the simpler situation described by \rf{const} and \rf{trace}, we find~:
\bea
{\cal L}_{e^2p^2} &=&
F^2\ \big\{
\,k_1\,\langle\,d^{\mu} U^+ d_{\mu} U\,\rangle
\,\langle\,Q^2\,\rangle
\nonumber\\
&&\qquad
+k_2\,\langle\,d^{\mu}U^+d_{\mu}U\,\rangle
\langle\,QUQU^+\,\rangle
\nonumber\\
&&\qquad
+k_3\,
\big(\,\langle\,d^{\mu}U^+QU\,\rangle\,
\langle\,d_{\mu}U^+QU\,\rangle\,+
\,\langle\,d^{\mu}UQU^+\,\rangle\,
\langle\,d_{\mu}UQU^+\,\rangle\,\big)
\nonumber\\
&&\qquad
+k_4\,\langle\,d^{\mu}U^+QU\,\rangle\,
\langle\,d_{\mu}UQU^+\,\rangle
\nonumber\\
&&\qquad
+k_5\,\langle\,\chi^+U+U^+\chi\,\rangle\,\langle\,Q^2\,\rangle
\nonumber\\
&&\qquad
+k_6\,\langle\,\chi^+U+U^+\chi\,\rangle\,
\langle\,QUQU^+\,\rangle
\nonumber\\
&&\qquad
+k_7\,\langle\,(\chi U^+ + U\chi^+ )Q+
(\chi^+U+U^+\chi )Q\,\rangle\,\langle\,Q\,\rangle
\nonumber\\
&&\qquad
+k_8\,\langle\,(\chi U^+ - U\chi^+ )QUQU^+
+(\chi^+U-U^+\chi )QU^+QU\,\rangle
\nonumber\\
&&\qquad
+k_9\,\langle\,d_{\mu}U^+[(c_R^{\mu}Q),Q]U
+d_{\mu}U[(c_L^{\mu}Q),Q]U^+\,\rangle
\nonumber\\
&&\qquad
+k_{10}\,\langle\,(c_R^{\mu}Q)U(c_{L{\mu}}Q)U^+\,\rangle
\nonumber\\
&&\qquad
+k_{11}\,\langle\,(c_RQ)\cdot(c_RQ)+(c_LQ)\cdot(c_LQ)\,\rangle\,\big\}\ ,
\lbl{ctr1}
\ena
and
\eq
{\cal L}_{e^4} \ =\ F^4\ \big\{\,
k_{12}\,
\langle\,Q^2\,\rangle^2
\,+k_{13}\,
\,\langle\,QUQU^+\,\rangle\,\langle\,Q^2\,\rangle
\,+k_{14}\,
\,\langle\,QUQU^+\rangle^2\,\big\}\ .
\lbl{ctr2}
\en
In the large $N_C$ limit, the constants $k_i$ are of order $O(1)$ for 
$i=1,\,...11$, and of order $O(1/N_C)$ for $i=12,13,14$. One combination of these constants is Zweig rule suppressed,
\eq
k_5+k_6-\frac{4}{5}k_7\ \sim\ O(1/N_C)\ .
\en
The renormalization of the low energy constants of ${\cal L}^{(4)}$ is given as
\bea
l_i &=\ l_i^r(\mu)+\gamma_i\lambda\quad &i=1,\dots\ 7\ ,\nonumber\\
h_i &=\ h_i^r(\mu)+\delta_i\lambda\quad &i=1,\,2,\,3\ ,\nonumber\\
k_i &=\ k_i^r(\mu)+\sigma_i\lambda\quad &i=1,\dots\ 14\ ,\lbl{ren}
\ena
with
\eq
\lambda ={{\mu^{d-4}}\over{16\pi^2}}\,\bigg(\,
{1\over{d-4}}-{1\over 2}[\ln 4\pi + \Gamma^{\prime}(1) + 1]\,\bigg)\ .
\en
The coefficients $\gamma_i$ and $\delta_i$ were computed in \cite{GL}, and can be read off from Eq. \rf{div}, which in addition yields 
\eq
\begin{array}{lll}
\sigma_1={\textstyle{-{27\over20}}}-{\textstyle{{1\over5}}}Z,
&\sigma_2=2Z,
&\sigma_3={\textstyle{-{3\over4}}},
\nonumber\\
\sigma_4=2Z,
&\sigma_5={\textstyle{-{1\over4}}}-{\textstyle{{1\over5}}}Z,
&\sigma_6={\textstyle{{1\over4}}}+2Z,
\nonumber\\
\sigma_7=0,
&\sigma_8={\textstyle{{1\over8}}}-Z,
&\sigma_9={\textstyle{{1\over4}}},
\nonumber\\
\sigma_{10}=0,
&\sigma_{11}=0,
&
\nonumber\\
\sigma_{12}={\textstyle{{3\over2}}}-{\textstyle{{12\over5}}}Z+
{\textstyle{{84\over25}}}Z^2,
&\sigma_{13}=-3-{\textstyle{{3\over5}}}Z-
{\textstyle{{12\over5}}}Z^2,
&\sigma_{14}={\textstyle{{3\over2}}}+3Z+12Z^2\ .
\end{array}\lbl{beta}
\en

Up to a reshuffling of the indices\footnote{We have numbered our $k_i$'s such as to make the correspondance with the SU(3) counterterms $K_i$ of \cite{res,neu} easier.} of the low energy 
constants $k_i$, the list \rf{ctr1} and \rf{ctr2} coincides with the one given by 
the authors of Ref. 
\cite{ulf}, {\it except} for the counterterm multiplied by $k_7$, which they have omitted. This term, 
however, cannot be rewritten as a linear combination of the remaining 
structures that appear in ${\cal L}_{e^2p^2}$, and must therefore be included, 
even if $k_7$ turns out not to be renormalized. In order to illustrate this 
point, we consider isospin breaking in the light quark 
condensates, which appears only at next-to-leading order, but receives no contributions from the pion loops,
\eq
<\,\Omega\,\vert\,{\bar u}u\,-\,{\bar d}d\,\vert\,\Omega\,>\ =
\ 4(m_d-m_u)B^2h_3+{8\over 3}e^2F^2Bk_7\ .
\en
For $e=0$, the result involves only the ``high energy'' 
constant $h_3$ \cite{GL}, which is a convention dependent short distance subtraction. 
In the chiral limit (or, to that order, for $m_u=m_d$) and with $e\neq 0$, the isospin violation in the 
condensates is given solely by the constant $k_7$. Both $h_3$ and $k_7$ are finite, $\delta_3=\sigma_7=0$, but, in contrast to 
the former, $k_7$ is an order parameter for the spontaneous breakdown of the 
SU(2)$\times$SU(2) chiral symmetry.

As far as the 
remaining $k_i$'s are concerned, Ref. \cite{ulf} also 
shows some 
discrepancies with our values of the renormalization factors $\sigma_i$. In a subsequent 
version \cite{ulf2}, its authors  brought their findings 
into agreement with our results \rf{beta}. Finally, the divergent part \rf{div} contains a renormalization of the Maxwell Lagrangian of the photon, which was not mentioned in \cite{ulf,ulf2}.

As a second application, we also give the expressions of the pion masses at next-to-leading order,

\bea
\Mpin &=& 2\mhat B\,\bigg\{\,
1+{{2\Mpin}\over{F^2}}l_3^r(\mu)+e^2{\cal K}_{\pi^0}^r(\mu)
+{{\Mpic}\over{16\pi^2F^2}}\,\ln\left({{\Mpic}\over{\mu^2}}\right)-
{{\Mpin}\over{32\pi^2F^2}}\,\ln\left({{\Mpin}\over{\mu^2}}\right)
\,\bigg\}
\nonumber\\
&&-2{{B^2}\over{F^2}}(m_d-m_u)^2\,l_7-{8\over3}B(m_d-m_u)e^2\,k_7
\nonumber\\
\nonumber\\
\Mpic &=& 2\mhat B\,\bigg\{\,
1+{{e^2}\over{4\pi^2}}+{{2\Mpin}\over{F^2}}l_3^r(\mu)+
e^2{\cal K}_{\pi^{\pm}}^r(\mu)
+{{\Mpin}\over{32\pi^2F^2}}\,\ln\left({{\Mpin}\over{\mu^2}}\right)
\,\bigg\}
\nonumber\\
&&+2e^2F^2\,\bigg\{\,Z(1+
{{e^2}\over{4\pi^2}})+e^2{\cal K}_{\pi^{\pm}}'^r(\mu)
-(3+4Z){{\Mpic}\over{32\pi^2F^2}}\,\ln\left({{\Mpic}\over{\mu^2}}\right)
\,\bigg\}
\nonumber\\
&&-{8\over3}B(m_d-m_u)e^2\,k_7\lbl{mass}
\ena
with
\bea
{\cal K}_{\pi^0}^r &=&
-{20\over9}\big[\,k_1^r+k_2^r-{\textstyle{9\over{10}}}(2k_3^r-k_4^r)-
k_5^r-k_6^r-{\textstyle{1\over5}}k_7\,\big]
\nonumber\\
{\cal K}_{\pi^{\pm}}^r &=&
-{20\over9}\big[\,k_1^r+k_2^r-
k_5^r-{\textstyle{1\over{5}}}(23k_6^r+k_7+18k_8^r)\,\big]
\nonumber\\
{\cal K}_{\pi^{\pm}}'^r &=&
-{10\over9}\,\big[\,2Z(k_1^r+k_2^r)-{\textstyle{1\over2}}k_{13}^r-
k_{14}^r\,\big]\ .
\ena
As expected, these expressions turn out to be independent of the choice of the renormalization scale $\mu$, which provides a non trivial check of the relations \rf{beta}. 
The low energy constants $l_7$ and $k_7$ induce corrections of the type $(m_d-m_u)^2$ and $(m_d-m_u)e^2$, respectively. While the latter do not contribute to the difference $\Mpic -\Mpin$, the former have been estimated in \cite{GLrep,GL} and were found to be negligible as compared to the leading order electromagnetic mass difference \rf{leadmass}. If, on the basis of naive dimensional analysis, we assume an upper bound on the various constants $k_i^r(\mu)$ at the scale $\mu=M_{\rho}$,
\eq
\vert\,k_i^r(M_{\rho})\,\vert\ \lapprox\ {1\over{16\pi^2}}\ ,\lbl{est}
\en
we obtain
\bea
e^2\,\vert\,{\cal K}_{\pi^0}^r(M_{\rho})\,\vert &\lapprox& 0.9\times 10^{-2}
\nonumber\\
e^2\,\vert\,{\cal K}_{\pi^{\pm}}^r(M_{\rho})\,\vert &\lapprox& 1.5\times 10^{-2}
\nonumber\\
e^2\,\vert\,{\cal K}_{\pi^{\pm}}'^r(M_{\rho})\,\vert &\lapprox& 0.3\times 10^{-2}\ ,
\ena
so that the corrections of order $O(e^2\mhat)$ and $O(e^4)$ to the pion masses
turn out to be numerically small. They are, for instance, comparable to the error on the $O({\mhat}^2)$ corrections generated by the uncertainty assigned to ${\bar l}_3$ in \cite{GL}, {\it viz.} ${\bar l}_3=2.9\pm 2.4$ or $(2\Mpin/F^2)\,l_3^r(M_{\rho})=(0.4\pm 1.6)\times 10^{-2}$. More elaborate statements require more information on the values of the constants $k_i$. In the present paper, since we are mainly interested in assessing the expected order of magnitude of electromagnetic corrections, we shall be content with the cruder estimate \rf{est}.

Before we turn to $\pi -\pi$ scattering, we should like to stress that the result \rf{beta} only holds in the Feynman gauge $a=1$. In general, the low energy constants $k_i$, and thus their divergences, depend on the choice of gauge. This gauge dependence has been displayed explicitly for some of the SU(3) counterterms $K_i$ of \cite{res,neu} in Ref. \cite{bachir}, but a systematic 
investigation has not been undertaken so far. Such a calculation would, for instance, be useful to check that the expressions for physical quantities, like the pion masses above, are not only scale independent, but indeed also 
gauge invariant.

\indent

\sect{The One Loop $\pi-\pi$ Scattering Amplitudes}
\setcounter{equation}{0}

In this Section, we compute the amplitudes for $\pi^0\pi^0\rightarrow\pi^0\pi^0$ and for $\pi^+\pi^-\rightarrow\pi^0\pi^0$ scattering at next-to-leading order in the presence of electromagnetic interactions and for $m_u=m_d$. The amplitudes involving charged pions only are however not considered here. 

For the process $\pi^0\pi^0\rightarrow\pi^0\pi^0$, the only differences with the calculation in the absence of electromagnetic interactions \cite{GL} consist in taking into account the contributions  from the counterterms of ${\cal L}_{e^2p^2}$ (those of 
${\cal L}_{e^4}$ do not contribute here) and in keeping track of the masses of the pions in the internal lines of the loop graphs. The amplitude is then given by \cite{ulf}
\bea
A^{00;00}(s,t,u) &=&
{{\Mpin}\over{F^2}}\,\bigg\{\,
1\,-\,{{\Mpin}\over{32\pi^2 F^2}}\,(3{\bar l}_3+1)\,+\,
{{\Mpin +2\Mpic}\over{16\pi^2 F^2}}\,L_{\pi}
\,+\,{{e^2}\over{32\pi^2}}\,{\cal K}^{00}\,\bigg\}
\nonumber\\
&&
+{1\over{48\pi^2F^4}}\,({\bar l}_1+2{\bar l}_2 -3L_{\pi}-3)\,\big[\,
(s-2\Mpin)^2+(t-2\Mpin)^2+(u-2\Mpin)^2\,\big]
\nonumber\\
&&
+{1\over{F^4}}\,\big[
(s-\Mpin)^2\Jpm(s)+(t-\Mpin)^2\Jpm(t)+(u-\Mpin)^2\Jpm(u)\,\big]
\nonumber\\
&&
+{{M_{\pi^0}^4}\over{2F^4}}\,\big[\,
\Joo(s) + \Joo(t) + \Joo(u)\,\big]\ ,\lbl{pi0}
\ena
where we have set
\eq
{\cal K}^{00}\ =\ (3+{{4Z}\over 9})\kbar_1-{{40Z}\over 9}\kbar_2-3\kbar_3
-4Z\kbar_4\ ,
\en
with
\eq
k_i^r(\mu)\ =\ {{\sigma_i}\over{32\pi^2}}\,\left(\,{\bar k}_i
+\ln\,{{\Mpin}\over{\mu^2}}\,\right)\ ,\ \ i\neq7\ ,
\en
and
\eq
 L_{\pi}\ =\ \ln\,{{\Mpic}\over{\Mpin}}\ .
\en
The loop functions ${\bar J}_{PQ}$ are as defined in \cite{GL2} (see also the Appendix). The subscript identifies the charges, and therefore the masses, of the two pions in the loop. As observed in \cite{ulf},  the contribution of the electromagnetic counterterms can be eliminated upon replacing $F$ by the neutral pion decay constant $F_{\pi^0}$\footnote{The coefficient of $L_{\pi}$ differs by a factor of two from the corresponding expression in \cite{ulf}.}, 
\eq
F_{\pi^0}\ =\ F\,\bigg\{\,
1\,-\,{{e^2}\over{32\pi^2}}\,\big[\,
(3+{{4Z}\over 9})\kbar_1-{{40Z}\over 9}\kbar_2-3\kbar_3
-4Z\kbar_4\,\big]\,+\,
{{\Mpin}\over{16\pi^2 F^2}}\,{\bar l}_4
\,-\,{{\Mpic}\over{16\pi^2 F^2}}\,L_{\pi}\,\bigg\}\ ,
\en
which describes the coupling of $\pi^0$ to the axial current,
\eq
<\Omega\,\vert\,{1\over2}({\bar u}\gamma_{\mu}\gamma_5u-
{\bar d}\gamma_{\mu}\gamma_5d)(0)\,\vert\,\pi^0(p)>\ =
\ ip_{\mu}\,F_{\pi^0}\ ,
\en
for $m_u=m_d$. 
However, since there is no accurate determination of this matrix element \cite{pdg}, this substitution is of little avail in practice. Instead, we shall use the charged pion decay constant $\Fpi$ in the absence of electromagnetism, defined as
\eq
<\Omega\,\vert\,({\bar u}\gamma_{\mu}\gamma_5d)(0)\,\vert\,
\pi^-(p)>\vert_{e=0}\ =
\ i\,\sqrt{2}\,p_{\mu}\,\Fpi\ ,
\en
with the value $\Fpi = 92.4$ MeV \cite{hol,mar,fink} as extracted from the $\pi_{\ell 2(\gamma)}$ decay rate\footnote{This extraction itself is not without ambiguities. In order to fix these, a more extended framework, allowing for a full treatment of radiative corrections in the semileptonic decays of the pseudoscalar mesons, is required.}, and the relation to $F$ given by \cite{GL}
\eq
\Fpi\ =\ F\,\bigg\{\,1\ +\ {{\Mpin}\over{16\pi^2 F^2}}\,{\bar l}_4\,\bigg\}\ .
\en

In the case of the process $\pi^+(p_1)\pi^-(p_2)\to\pi^0(k_1)\pi^0(k_2)$, one  has to consider in addition loops with virtual photons. These affect the wave function renormalization of the charged pions and introduce a long range component into the scattering amplitude through the vertex correction graph of Fig. 1. With the Condon-Shortley phase convention, the amplitude reads
\eq
A^{+-;00}(s,t,u)\ =
\ -\ {{s-\Mpin}\over{F^2}}\ -\ B(s,t,u)\ -\ C(s,t,u)\ , 
\en

\vspace{7mm}

\centerline{\psfig{figure=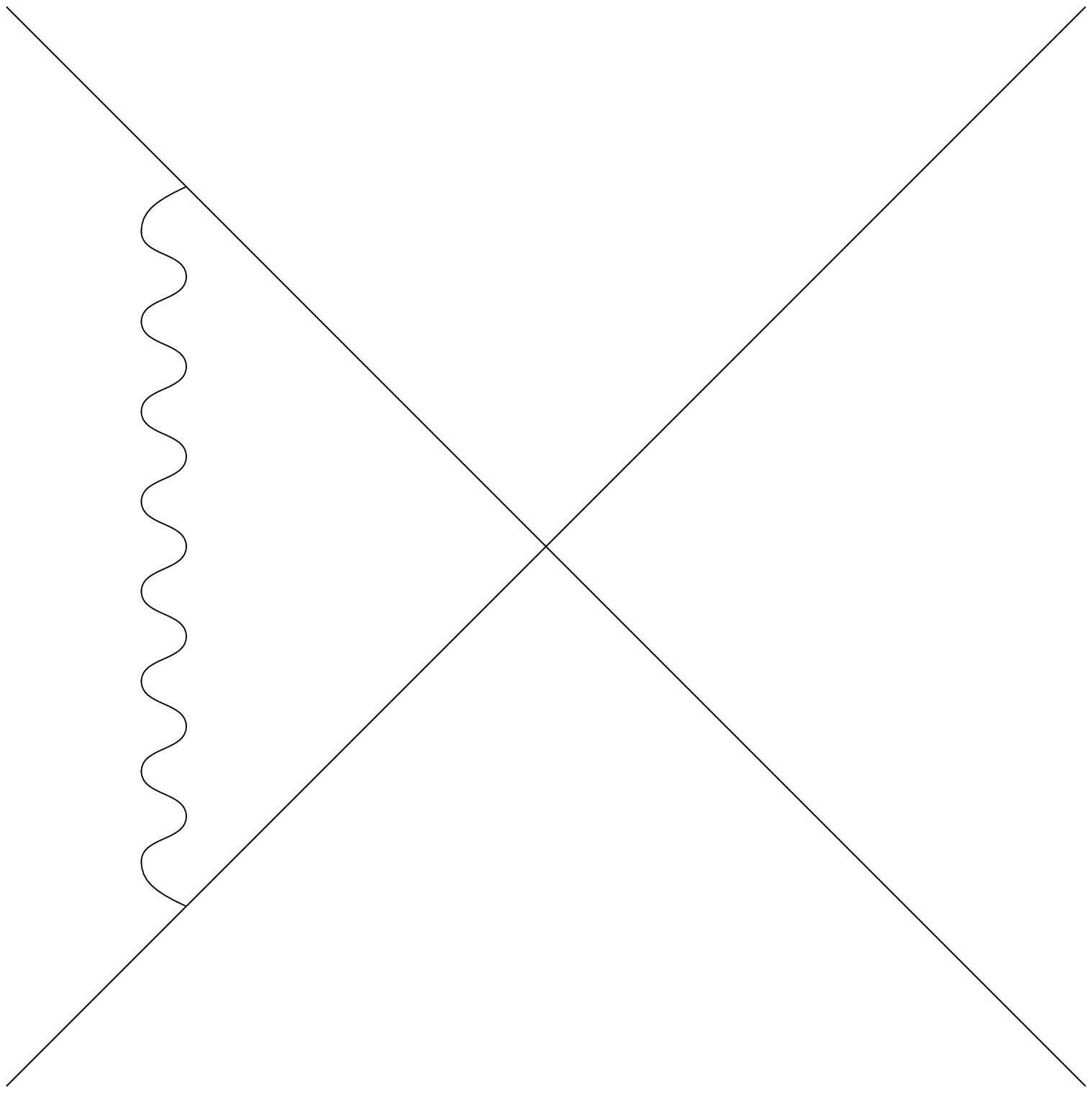,height=2.5cm,angle=-90}}
{\bf Figure 1} : The one photon exchange electromagnetic correction to the strong vertex.


where $B(s,t,u)$ represents the unitarity corrections 
\bea
B(s,t,u) &=&
{{s-\Mpin}\over {F^4}}\,\bigg\{\,{\Mpin \over 2}\,\Joo(s)
+ \bigg(\,{s \over 2}+2\Delta_{\pi} - e^2F^2\,\bigg)\,\Jpm(s)
\nonumber\\
&&\qquad\qquad
-e^2F^2\,(2s-4\Mpic)\,G_{+-\gamma}(s)\,\bigg\}
\nonumber\\
&&
+{1\over{12F^4}}\,\bigg\{\,3\,\bigg[\,
t-2\Mpic+{{\Delta_{\pi}^2}\over t}\,\bigg]^2
\nonumber\\
&&\qquad\qquad
+{{\lambda(t,\Mpic,\Mpin)}\over{t^2}}\,\bigg[\,
\Delta_{\pi}^2+t(s-u)\,\bigg]\,\bigg\}\,{\bar J}_{+0}(t)
\nonumber\\
&&
+{1\over{12F^4}}\,\bigg\{\,3\,\bigg[\,
u-2\Mpic+{{\Delta_{\pi}^2}\over u}\,\bigg]^2
\nonumber\\
&&\qquad\qquad
+{{\lambda(u,\Mpic,\Mpin)}\over{u^2}}\,\bigg[\,
\Delta_{\pi}^2+u(s-t)\,\bigg]\,\bigg\}\,{\bar J}_{+0}(u)\ ,
\ena
whereas the contributions from $O(p^4)$ tree and tadpole graphs have 
been collected in $C(s,t,u)$ ($\Sigma_{\pi}\ =\ \Mpic+\Mpin$),
\bea
C(s,t,u) &=&
{{s-\Mpin}\over{32\pi^2F^4}}\,\bigg\{\,
-{{\Sigma_\pi}\over 3}-4\Delta_\pi
-{{L_\pi}\over{\Delta_{\pi}}}\,
\bigg(4M_{\pi^\pm}^4 - 7\Mpin\Mpic + 5M_{\pi^0}^4\bigg)
\nonumber\\
&&
\qquad\qquad
+e^2F^2\,\bigg[\,2+(3+{{4Z}\over 9})\kbar_1+{{32Z}\over 9}\kbar_2+3\kbar_3
+4Z\kbar_4-6L_{\pi}+4 \ln {\Mpic\over{m_\gamma^2}}\,\bigg]\,\bigg\}
\nonumber\\
&&
-{{\Mpin}\over{32\pi^2F^4}}\,\bigg\{\,
{{\Mpic}\over 3}-{{10\Mpin}\over 9}
-{{L_\pi}\over{\Delta_{\pi}}}\,\Sigma_\pi
\bigg(2\Mpic - \Mpin\bigg) + \Mpin {\bar l}_3
\nonumber\\
&&
\qquad\qquad
+e^2F^2\,\bigg[\,8Z\kbar_2+3\kbar_3+4Z\kbar_4
-2(1+8Z)\kbar_6-(1-8Z)\kbar_8\,\bigg]
\nonumber\\
&&
+{{M_{\pi^{\pm}}^4}\over{24\pi^2F^4}}\,({1\over 3}+L_{\pi})
-{{\Delta_{\pi}}\over{96\pi^2F^4}}\,\bigg(\,{1\over t}+{1\over u}\,\bigg)\,
(\Sigma_{\pi}\Delta_{\pi} - 2\Mpin\Mpic L_{\pi})
\nonumber\\
&&
-{1\over{48\pi^2F^4}}\,\bigg\{\,
{1\over 6}\,(\,11s^2-t^2-u^2\,)
+ {{L_\pi}\over{\Delta_{\pi}}}\,\bigg[\,
\bigg(\,\Mpic - {3\over 2}\,\Mpin\,\bigg)\,s^2+\Mpic\,(t^2+u^2)\,\bigg]\,
\bigg\}
\nonumber
\\
&&
+{1\over{48\pi^2F^4}}\,{\bar l}_1\,(s-2\Mpin)(s-2\Mpic)
\nonumber\\
&&
+{1\over{48\pi^2F^4}}\,{\bar l}_2\,\bigg[\,
(t-\Sigma_{\pi})^2+(u-\Sigma_{\pi})^2\,\bigg]\ .
\ena

The loop function $G_{+-\gamma}(s)$ involves three propagators, and comes from the Feynman graph of Fig. 1, 
\eq
G_{+-\gamma}(s)\ =\ -i\,\int\,{{d^4q}\over{(2\pi)^4}}\,
{1\over{(q^2-m_{\gamma}^2)(q^2-2q\cdot p_1)(q^2+2q\cdot p_2)}}\ .\lbl{Gpmg}
\en 
While being ultraviolet finite, this function develops an infrared singularity as the photon mass $m_{\gamma}$ is sent to 
zero. In order to display this singularity, we express the function $G_{+-\gamma}(s)$ in terms of the dilogarithm or Spence function ${\rm Li}_2(x)$ (see the Appendix). For $s>4\Mpic$, and for $m_{\gamma}\to 0$, it is given as follows~:
\bea
G_{+-\gamma}(s) &=& -\,{1\over{32\pi^2s\sigma}}\,\bigg\{\,
4\,{\rm Li}_2\left({{1-\sigma}\over{1+\sigma}}\right)
+{{4\pi^2}\over3}+\ln^2\left({{1-\sigma}\over{1+\sigma}}\right)
\nonumber\\
&&\quad
+2\,\bigg[\,\ln\left({s\over{\Mpic}}\right)
-\ln\left({m_{\gamma}^2\over{\Mpic}}\right)
+2\ln(\sigma)\,\bigg]\,
\left[\,\ln\left({{1-\sigma}\over{1+\sigma}}\right)+i\pi\,\right]\,\bigg\}
\ ,
\lbl{Gpmg2}
\ena
where
\eq
\sigma\ =\ \sqrt{1-{{4\Mpic}\over s}}\ .
\en
Both the real and imaginary parts of $G_{+-\gamma}(s)$ diverge at threshold. Another infrared singular piece, coming from the wave function renormalization of the charged pions, appears in $C(s,t,u)$. 
As is well known, in order to obtain an infrared finite cross section, one has to consider the processes with emissions of soft photons. This question will be addressed in the next Section. For the moment, we furthermore 
notice that both $B(s,t,u)$ and $C(s,t,u)$ contain terms which involve inverse powers of the Mandelstam variables $t$ and $u$, and whose coefficients vanish in the limit where the neutral and charged pion masses become equal. In fact, the coefficient of these terms in $C(s,t,u)$ is rather small~: Expanding the logarithm $L_{\pi}$ in powers of $\Delta_{\pi}$, one has
\eq
\Sigma_{\pi}\Delta_{\pi} - 2\Mpin\Mpic L_{\pi}\ =
\ {{M_{\pi^0}^4}\over3}\cdot \left(\,
{{\Delta_{\pi}}\over{\Mpin}}\,\right)^3\,\left[\,1\ +
\ O\left(\,
{{\Delta_{\pi}}\over{\Mpin}}\,\right)\,\right]\ ,
\en
which yields a negligible contribution even near threshold, where 
$t\sim u\sim -\Delta_{\pi}$. Keeping only terms which are at most of order 
$\Delta_{\pi}^2/M_{\pi^0}^4$, the expression of 
$C(s,t,u)$ simplifies somewhat
\bea
C(s,t,u)&=&
{{s-\Mpin}\over{32\pi^2F^4}}\,\left(\,
-4\Mpin -7\Delta_\pi -{{23\over6}}\,{{\Delta_\pi^2}\over {\Mpin}} 
+ e^2F^2({\cal K}^{\pm 0}_1 + 2)+4e^2F^2 \ln {\Mpic\over{m_\gamma^2}}\,\right)
\nonumber\\
&&
-{{\Mpin}\over{32\pi^2F^4}}\,\left(\,
  \Mpin\,({\bar l}_3-1) 
-3\,{\Delta_{\pi}}-{{29\over{18}}}\,{{\Delta_\pi^2}\over{\Mpin}}
+ e^2F^2{\cal K}^{\pm 0}_2\,\right)
\nonumber\\
&&
+{1\over{48\pi^2F^4}}\,\left(\,{\bar l}_1-{{4\over3}}-
{{5\over4}}\,{{\Delta_\pi}\over{\Mpin}}+
{{2\over3}}\,{{\Delta_\pi^2}\over{M_{\pi^0}^4}}\,\right)
\,(s-2\Mpin)(s-2\Mpic)
\nonumber\\
&&
+{1\over{48\pi^2F^4}}\,\left(\,{\bar l}_2-{{5\over6}}-
{{1\over2}}\,{{\Delta_\pi}\over{\Mpin}}+
{{1\over6}}\,{{\Delta_\pi^2}\over{M_{\pi^0}^4}}\,\right)
\,\bigg[\,(t-\Sigma_{\pi})^2+(u-\Sigma_{\pi})^2\,\bigg]\ ,
\ena
with
\bea 
{\cal K}^{\pm 0}_1 &=&  
(3+{{4Z}\over 9})\kbar_1+{{32Z}\over 9}\kbar_2+3\kbar_3
+4Z\kbar_4-6L_{\pi}\ ,
\nonumber\\
{\cal K}^{\pm 0}_2 &=& 
8Z\kbar_2+3\kbar_3+4Z\kbar_4
-2(1+8Z)\kbar_6-(1-8Z)\kbar_8\ .
\ena
In the sequel, we shall always neglect terms which are smaller than $\Delta_{\pi}^2/M_{\pi^0}^4$ or than $e^2\,\Delta_{\pi}/M_{\pi^0}^2$. 

As far as the representation \rf{iso} is concerned, if it were still correct at next-to-leading order, it would entail the following 
relation among the two amplitudes we have obtained above~:
\eq
-\,A^{00;00}(s,t,u)\ =\ A^{+-;00}(s,t,u)\,+\, A^{+-;00}(t,u,s)
\,+\, A^{+-;00}(u,s,t)\ ,\lbl{isoNLO}
\en
once the one photon exchange and infrared singular contributions have been subtracted from the amplitude $A^{+-;00}(s,t,u)$.
Even without figuring out how to perform this step in an unambiguous way at the present stage, it is clear that such a relation cannot hold~: Not only are the unitarity corrections due to two pion intermediate states different from what the relation \rf{isoNLO} would require, but also the contributions of the electromagnetic counterterms at order $O(e^2\mhat)$ do not satisfy this relation. If one considers in addition the scattering amplitudes involving four charged pions, one can easily see, even without doing a full calculation, that they will involve the constant $k_{14}$ of ${\cal L}_{e^4}$, which does not contribute to the  amplitudes involving neutral pions or to the masses. Therefore, we conclude that the representation \rf{iso} in terms of a single amplitude $A(s\vert t,u)$ is violated by next-to-leading corrections both of order $O(e^2p^2)$ 
and of order $O(e^4)$. The meaning to be assigned to the amplitude defined by Eqs. (22), (23) and (24) of Ref. \cite{ulf} thus remains unclear to us. 

More than twenty years ago, Roig and Swift \cite{rs} had already investigated electromagnetic corrections of order $O(e^2p^2)$ to the $\pi-\pi$ scattering 
amplitudes. Their approach was not systematic from the point of view adopted in the present article~: the pion loops and the pion mass difference were not taken into account. Thus, up to the Born terms, only the photon loop induced wave function renormalization of charged pions and vertex correction graphs of the type shown in  Fig. 1 were included in their calculation. Their claim, however, was that the resulting amplitudes were all finite. We have checked that this is not the case for $\pi^+\pi^-\to\pi^0\pi^0$ (the same conclusion was reached independently in Ref. \cite{wic}). Working within the framework adopted by Roig and Swift, we obtain (now in an arbitrary gauge $a$)
\bea
A^{+-;00}_{\rm RS}(s,t,u) &=&
-\,{{s-\Mpic}\over{F^2}}\,\bigg\{\, 1 + {{e^2}\over{16\pi^2}}
-2e^2(s-\Mpic)G_{+-\gamma}(s)-e^2\Jpm(s)
\nonumber\\
&&\qquad
-{{e^2}\over{16\pi^2}}\,\bigg[\,
3\ln\left( {{\Mpic}\over{\mu^2}}\right) 
-2\ln\left( {{\Mpic}\over{m_{\gamma}^2}}\right)
\,\bigg]
\,-\,6e^2\lambda\,\bigg\}\ .\lbl{AmpRS}
\ena
Thus, although we end up with a gauge invariant result, it however diverges as $d\to 4$. This disease is cured by the counterterms $k_i$. We have also checked that the amplitudes involving four charged pions are divergent in the framework of Ref. \cite{rs}. Thus, omitting the contributions of the electromagnetic counterterms to the scattering amplitudes does not lead to a consistent result.

\indent

\sect{Scattering Lengths}
\setcounter{equation}{0}

Evaluating the amplitude \rf{pi0} at threshold, $s=4\Mpin$, $t=u=0$, we obtain the following expression for the scattering length,
\eq
a_0(00;00) \ =  
\ {{\Mpin}\over{32\pi \Fpi^2}}\,\bigg\{\,
1\,+\,{{\Mpin}\over{32\pi^2 \Fpi^2}}\,
\bigg[\,8{\bar l}_1+16{\bar l}_2-3{\bar l}_3+4{\bar l}_4+13
\,+\,
18\,{{\Delta_{\pi}}\over{\Mpin}}\,+\,{{\Delta_{\pi}^2}\over{M_{\pi^0}^4}}
\,\bigg]\,+\,{{e^2}\over{32\pi^2}}\,{\cal K}^{00}
\,\bigg\}\ ,\lbl{a0000}
\en
where we have neglected terms of order $\Delta_{\pi}^3/M_{\pi^0}^6$ and beyond.  At next-to-leading order, $M_{\pi}^2$ coincides neither with $\Mpic$ nor with $\Mpin$ for $e\neq 0$. We shall define the correction $\Delta a_0(00;00)$ as in Section 1, Eq. \rf{adef}, but now with \cite{GL}
\bea
(a_0^0)_{\rm str} &=&{{7\Mpic}\over{32\pi\Fpi^2}}\,\bigg\{\,
1+{5\over{84\pi^2}}\,{{\Mpic}\over{\Fpi^2}}\,
\bigg[\,{\bar l}_1+2{\bar l}_2-{3\over 8}{\bar l}_3+{{21}\over{10}}{\bar l}_4
+{21\over 8}\,\bigg]\,\bigg\}\ =\ 0.20\pm 0.01
\\
(a_0^2)_{\rm str} &=&-\,{{\Mpic}\over{16\pi\Fpi^2}}\,\bigg\{\,
1-{1\over{12\pi^2}}\,{{\Mpic}\over{\Fpi^2}}\,
\bigg[\,{\bar l}_1+2{\bar l}_2-{3\over 8}{\bar l}_3-{{3}\over{2}}{\bar l}_4
+{3\over 8}\,\bigg]\,\bigg\}\ =\ -0.043\pm0.004 .\nonumber 
\ena
For the numerical evaluations, we use the central values of the low energy constants ${\bar l}_i$ that were given in \cite{GL}, ${\bar l}_1=-2.3\pm 3.7$, ${\bar l}_2=6.0\pm 1.3$, ${\bar l}_3=2.9\pm 2.4$, ${\bar l}_4=4.3\pm 0.9$.
With the estimate \rf{est}, we find
\bea
&&{{e^2\Fpi^2}\over{\Mpin}}\,{\cal K}^{00}\ =\ 1.0\pm 0.9\ ,
\nonumber\\
&&{{e^2\Fpi^2}\over{\Mpin}}\,{\cal K}^{\pm 0}_1\ =\ 1.8\pm 0.9\ ,
\nonumber\\
&&{{e^2\Fpi^2}\over{\Mpin}}\,{\cal K}^{\pm 0}_2\ =\ 0.5\pm 2.2\ .
\ena
and
\eq
\Delta a_0(00;00)\ =\ (-3.2\pm 0.1)\times 10^{-3}\ ,\lbl{num1}
\en
which corresponds to a relative variation of $-8\%$ as compared to the one loop value $a_0(00;00)\vert_{\rm str}$ $=-0.039$ obtained in the absence of electromagnetism. We see also that the contribution of the counterterms $k_i^r$ to the correction, which is reflected by the uncertainty in \rf{num1}, is very small, and would remain so even if the estimate \rf{est} turned out to be wrong by a factor of two or three.

As we have mentioned earlier, the amplitude for $\pi^+\pi^-\to\pi^0\pi^0$ contains an infrared divergent contribution generated by the long range electromagnetic interactions between the incoming charged pions and by their wave function renormalization. 
The threshold expansion of the real part of the above amplitude takes the following form ($q$ denotes the momentum of the charged pions in the center of mass frame)
\eq
Re A^{+-;00}(s,t,u) \ =
\ -{{4\Mpic-\Mpin}\over{\Fpi^2}}\cdot{{e^2}\over16}\cdot
{{M_{\pi^{\pm}}}\over{q}}+Re A^{+-;00}_{\rm thr}+O(q)\ ,
\en
with
\bea
Re\,A^{+-;00}_{\rm thr}
&=& 32\pi\,\left[\,-{1\over 3}(a_0^0)_{\rm str}\,+\,{1\over 3}(a_0^2)_{\rm str}\,\right]
\,-\,{{\Delta_{\pi}}\over{\Fpi^2}}\,
+\,{{e^2\Mpin}\over{32\pi^2\Fpi^2}}\,(\,30-3{\cal K}^{\pm 0}_1+
{\cal K}^{\pm 0}_2\,)
\nonumber\\
&&-{{\Delta_{\pi}}\over{48\pi^2\Fpi^4}}\,
\big[\,\Mpin(1+4{\bar l}_1+3{\bar l}_3-12{\bar l}_4)-
6\Fpi^2e^2(10-{\cal K}_1^{\pm 0})\,\big]
\nonumber\\
&&+{{\Delta_{\pi}^2}\over{480\pi^2\Fpi^4}}\,
\big[\,212-40{\bar l}_1-15{\bar l}_3+180{\bar l}_4\,\big]\ .
\ena
For the amplitude of the reaction $\pi^+\pi^0\rightarrow\pi^+\pi^0$, the corresponding expressions read
\eq
Re A^{+0;+0}(s,t,u) \ =
Re A^{+0;+0}_{\rm thr}+O(q)\ ,
\en
and
\bea
Re A^{+0;+0}_{\rm thr} &=& 16\pi\,(a_0^2)_{\rm str}
\,+\,{{\Delta_{\pi}}\over{\Fpi^2}}\,
-\,{{e^2\Mpin}\over{32\pi^2\Fpi^2}}\,(\,2+{\cal K}^{\pm 0}_1+
{\cal K}^{\pm 0}_2\,)
\nonumber\\
&&
-{{\Mpin\Delta_{\pi}}\over{48\pi^2\Fpi^4}}\,
\big[\,9+4{\bar l}_1+8{\bar l}_2-3{\bar l}_3-12{\bar l}_4\,\big]
\nonumber\\
&&
+{{\Delta_{\pi}^2}\over{288\pi^2\Fpi^4}}\,
\big[\,10-24{\bar l}_1-48{\bar l}_2+9{\bar l}_3+36{\bar l}_4\,\big]\ .
\ena
Numerically, we find
\bea
&&{1\over{32\pi}}\,Re A^{+-;00}_{\rm thr}-
\left[\,-{1\over 3}(a_0^0)_{\rm str}\,+\,{1\over 3}(a_0^2)_{\rm str}\,\right]
\ =\ (-1.2\pm 0.7)\times 10^{-3}\ ,
\nonumber\\
&&{1\over{32\pi}}\,Re A^{+0;+0}_{\rm thr}-
{1\over 2}(a_0^2)_{\rm str}
\ =\ (1.3\pm 0.4)\times 10^{-3}\ .
\ena
The error bars reflect the naive dimensional estimates \rf{est} of the contributions from the low energy constants $k_i^r(M_{\rho})$, which again are rather 
small.
Although the infrared singularity, which contributes only to the terms of order $O(q^2)$ or higher, affects neither the constant terms $Re\,A^{+-;00}_{\rm thr}$ and $Re\,A^{+0;+0}_{\rm thr}$ nor the long range part of the one photon exchange graph, scattering lengths ought to be defined from infrared finite observables. In the remaining part of this Section, we shall perform this last step in the case of the process $\pi^+\pi^-\rightarrow\pi^0\pi^0$. 

An infrared finite {\it cross section} is obtained if one takes into account the emission of soft photons with energies below the detector resolution. At the order we are working, it is enough to consider the emission of a single photon with the $O(p^2)$ $\pi^+\pi^-\pi^0\pi^0$ vertex taken from \rf{p2}. The corresponding amplitude reads
\eq
A^{+-;00\gamma}\ =\ {e\over{F^2}}\,\left[\,(k_1+k_2)^2-\Mpin\,\right]\,\epsilon^{\mu}(k_{\gamma})
\,\left\{\,{{(2p_1-k_{\gamma})_{\mu}}\over{m_{\gamma}^2-2p_1\cdot k_{\gamma}}}-
{{(2p_2-k_{\gamma})_{\mu}}\over{m_{\gamma}^2-2p_2\cdot k_{\gamma}}}\,\right
\} \ ,
\en
where $k_{\gamma}$ and $\epsilon_{\mu}(k_{\gamma})$ are the photon momentum 
and polarization, respectively.
The infrared finite cross section is then given by the sum
\eq
\sigma_{\rm tot}(s;{\Delta E})\ =\ \sigma^{+-;00}(s)\ +
\ \sigma^{+-;00\gamma}(s;{\Delta E})\ ,\lbl{tot}
\en
with
\eq
\sigma^{+-;00}(s)\ =\ {1\over{32\pi s}}\,\sqrt{{{s-4\Mpin}\over{s-4\Mpic}}}\,
\times\,{1\over2}\,\int\,d(\cos\theta)\,\left\vert\,A^{+-;00}\,\right\vert^2
\en
being the cross section without emitted photon, whereas the cross section with one emitted photon reads
\bea
&&\sigma^{+-;00\gamma}(s;{\Delta E}) \ =\ {\tilde\sigma}^{+-;00\gamma}
(s;{\Delta E})
\nonumber\\
&&
-\ {1\over{64\pi^3s}}\cdot{{e^2}\over{F^4}}\cdot
\sqrt{{{s-4\Mpin}\over{s-4\Mpic}}}\,(s-\Mpin)^2
\,\left[\,1+{{s-2\Mpic}\over{s\sigma}}\,
\ln\left({{1-\sigma}\over{1+\sigma}}\right)\,\right]\,
\ln\left({{\Delta E}\over{m_{\gamma}}}\right) .
\ena
The infrared finite piece ${\tilde\sigma}^{+-;00\gamma}(s;{\Delta E})$ involves an integration over the energy of the emitted photon, which, in the present case, can be done analytically, and reads
\bea
{\tilde\sigma}^{+-;00\gamma} &=& -{1\over{64\pi^3s}}\cdot{{e^2}\over{F^4}}\cdot
\sqrt{{{s-4\Mpin}\over{s-4\Mpic}}}\,
\bigg\{\,\bigg[\,1+{{s-2\Mpin}\over{s\sigma}}\,
\ln\left({{1-\sigma}\over{1+\sigma}}\right)\,\bigg]\,[F(s;\Delta E)-F(s;0)]
\nonumber\\
&&\qquad\qquad\qquad
+\,(s-\Mpin)^2\,\left[\,
\ln2+{1\over{2\sigma}}\,\ln\left({{1-\sigma}\over{1+\sigma}}\right)
\,\right]\,\bigg\}\ .\lbl{sgamma}
\ena
The function $F(s;\Delta E)$ is given in the Appendix.
The maximal energy $\Delta E$ of the undetected photon is limited by the detector resolution $\Delta E_{\rm det}$ or, for $s$ close to threshold, by the available phase space (in the present case, the maximal photon energy at threshold is $\Delta_{\pi}/M_{\pi^{\pm}}=9$ MeV),
\eq
\Delta E\ =\ {\rm min}\,
\big(\,{{s-4\Mpin}\over{2\sqrt{s}}}\, ,\, \Delta E_{\rm det}\,\big)\ .
\en
As seen from the above expressions, the infrared divergent pieces cancel in the sum \rf{tot}. 

Following Roig and Swift \cite{rs}, we may define the scattering length $a_0(+-;00)$ from the threshold expansion of the infrared finite cross section $\sigma_{\rm tot}$,
\eq
\sigma_{\rm tot} \ = 
\ {1\over{32\pi s}}\,\sqrt{{{s-4\Mpin}\over{s-4\Mpic}}}
\ \bigg\{\,-{{4\Mpic-\Mpin}\over{\Fpi^2}}\cdot{{e^2}\over16}\cdot
{{M_{\pi^{\pm}}}\over{q}}\,+\,32\pi a_0(+-;00)\,+\,O(q)\,\bigg\}^2\ .
\en
Notice that we have not expanded the ratio of the phase space by the flux factor in front of $\sigma_{\rm tot}$.
Up to terms which are of higher order in the chiral expansion, the scattering length thus defined reads
\eq
32\pi a_0(+-;00) \ = 
\ Re A^{+-;00}_{\rm thr}-{{e^2}\over{\pi^2\Fpi^2}}\,(1-\ln 2)\,
\left[\,{3\over 4}\Mpin - {1\over 2}\Delta_{\pi}\,\right]\ ,
\en
and does not depend on $\Delta E$ \cite{rs}. However, as compared to Eq. (5.6), it receives an additional contribution, the second term on the right hand side of (5.17), which is specific to the infrared finite observable $\sigma_{\rm tot}$ that we have considered here. Numerically, this represents only a tiny

\centerline{\psfig{figure=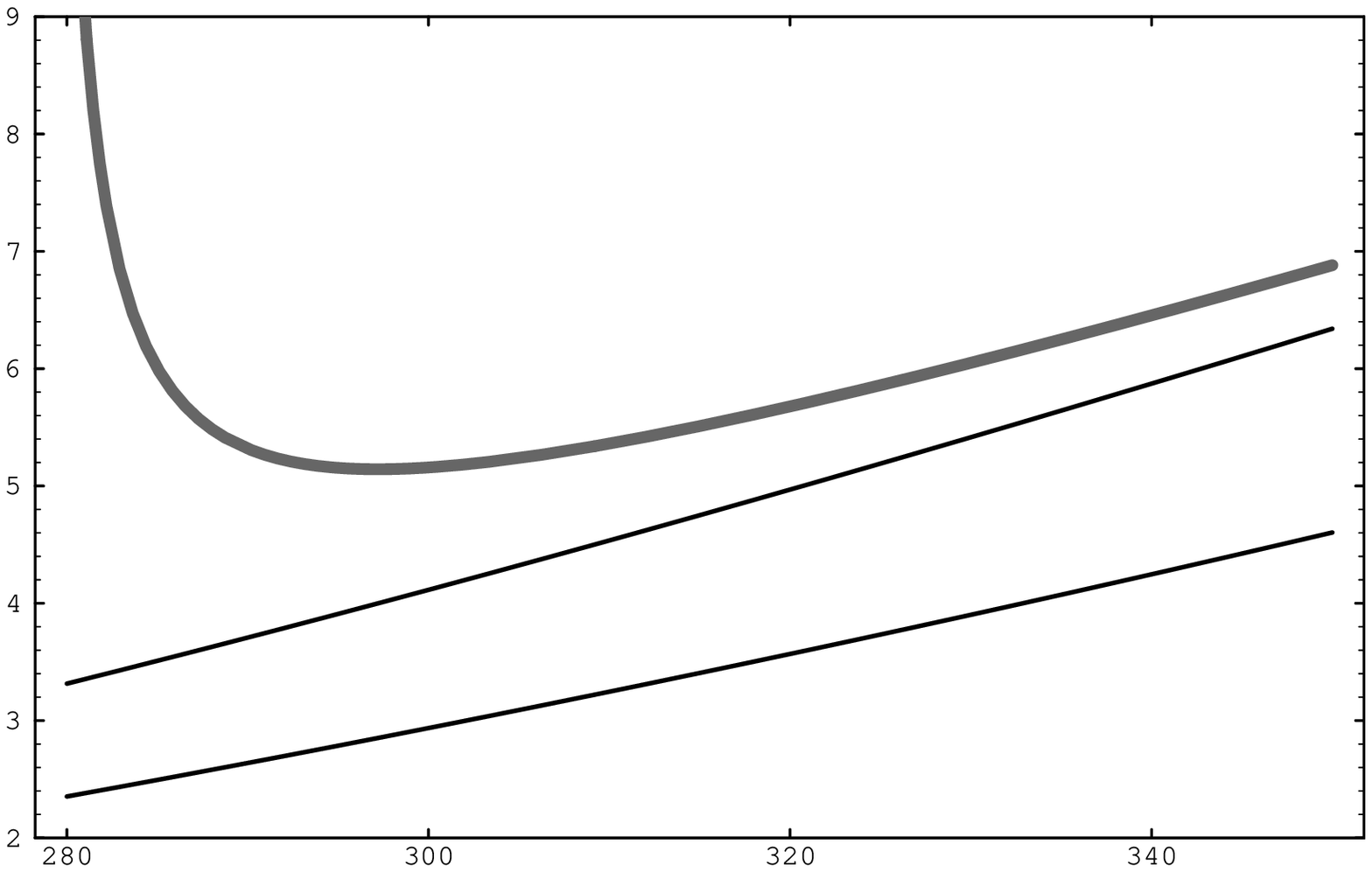,height=10.5cm}}

\noindent
{\bf Figure 2} : From bottom to top, the pure strong interaction cross sections for $\pi^+\pi^-\rightarrow\pi^0\pi^0$ at leading and at next-to-leading order, and the infrared finite radiative cross section $\sigma_{\rm tot}$ (in millibarn) as a function of the center of mass energy (in MeV).

\vspace{15mm}

\noindent
modification, and we obtain
\eq
\Delta a_0(+-;00)\ =\ (-\,1.2\pm 0.7)\times 10^{-3}\ ,\lbl{num2}
\en
corresponding to a relative variation between $-2.3\%$ and $-0.6\%$ as compared to the one loop value $a_0(+-;00)=-0.082$ for $e=0$.
Upon comparing Eqs. \rf{num1} and \rf{num2} with the values obtained at leading order $O(e^2)$ in Section 2, Eq. \rf{a0}, we conclude that the electromagnetic shift in the scattering length $a_0(00;00)$ including the $O(e^2p^2,e^4)$ contributions is twice as large as the leading order $O(e^2)$ effect, whereas the variation of $a_0(+-;00)$ is not strongly affected by next-to-leading order electromagnetic corrections.

Finally, we compare the total cross section $\sigma_{\rm tot}$ to the cross section for $\pi^+\pi^-\rightarrow\pi^0\pi^0$ scattering in the absence of electromagnetic interactions. The cross sections are shown in Fig. 2. We have taken a typical detector resolution of $\Delta E= 20$ MeV. The thickness of the upper curve shows the uncertainty induced by the estimate \rf{est} of the counterterms of ${\cal L}_{e^2p^2}$. The error coming from the uncertainties of the strong counterterms ${\bar l}_{1,2,3,4}$ are not shown. In Fig. 3, we show the ratio of $\sigma_{\rm tot}$ and of the one loop strong cross section with the leading order cross section for $e=0$.  As one approaches the threshold, the long range Coulomb contribution and the flux factor make the cross section diverge. Below 300 MeV, these two effects become more important than the strong one loop corrections themselves.

\centerline{\psfig{figure=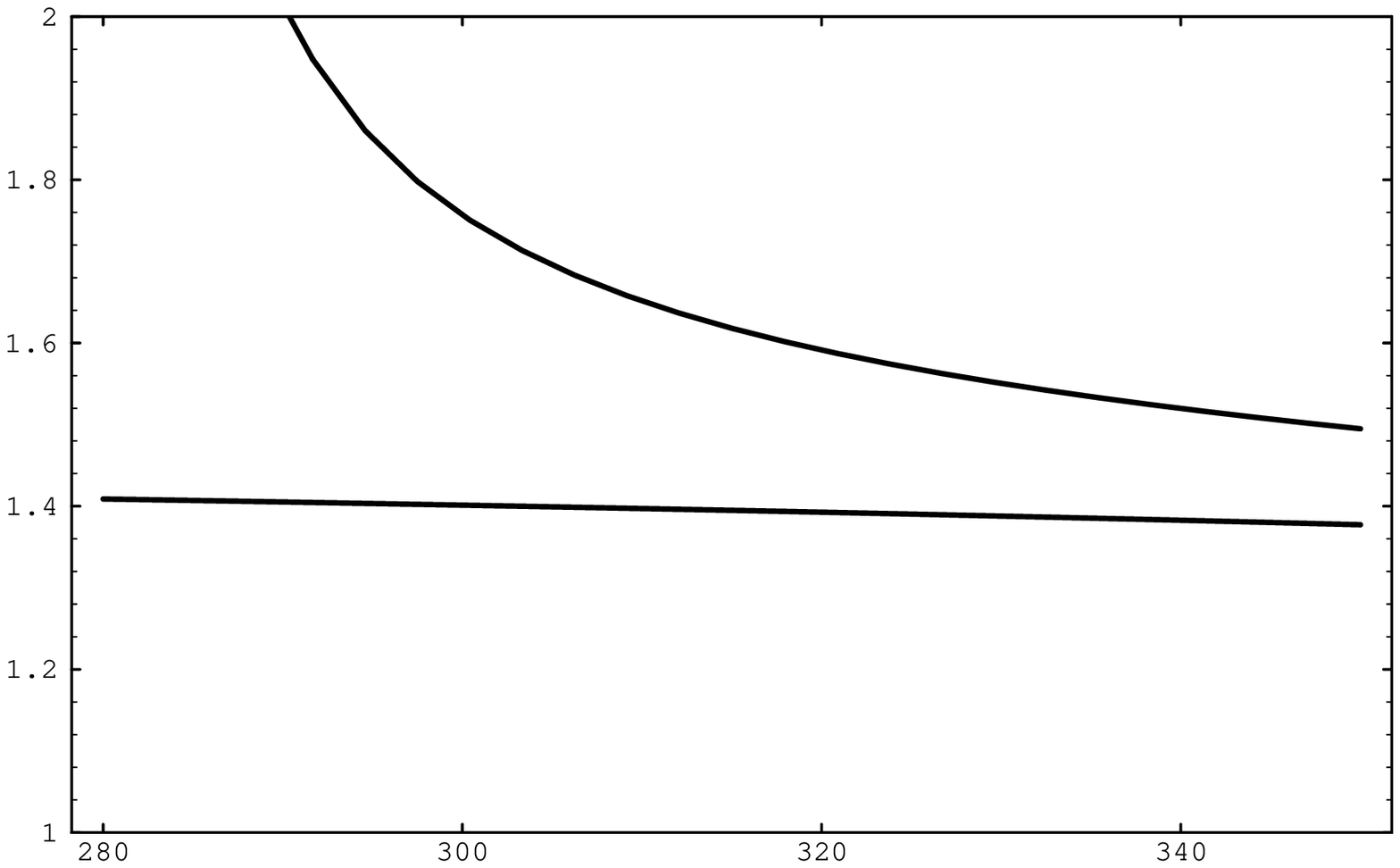,height=10.5cm}}

\noindent
{\bf Figure 3} : The relative corrections to the leading order cross section with only strong interaction corrections (lower curve), and with electromagnetic effects included at next-to-leading order (upper curve), as a function of the center of mass energy (in MeV).

\indent

\sect{Summary}
\setcounter{equation}{0}

\indent

In the present paper, electromagnetic corrections to low energy $\pi-\pi$ scattering were investigated at leading and next-to-leading order, within a systematic and consistent framework \cite{res,neu} which includes loops with 
virtual photons and combines the usual chiral expansion with the expansion in powers of the electric charge.

At leading order, the only direct effect of the electromagnetic interaction between quarks is the splitting between the masses of the charged and neutral pions. In particular, a single amplitude allows to describe all the different $\pi-\pi$ scattering channels at this order. All the S-wave scattering lengths for the channels involving at least one pair of charged pions can be expressed, as in the absence of electromagnetism, in terms of two suitaby redefined isospin scattering lengths $a_0^0$ and $a_0^2$, Eq. \rf{acorr}. Only the expression for $a_0(00;00)$ receives an additional contribution, which explicitly breaks isospin symmetry. We have found that numerically these corrections are comparable in magnitude to those induced at two loop level by the strong interactions alone, as evaluated in Refs. \cite{bcegs,gkms}.

In order to obtain the $\pi-\pi$ scattering amplitudes at next-to-leading order, we have computed the generating functional in the presence of virtual photons to one loop in the two flavour case. Its divergences (in the Feynman gauge) and the complete list of relevant counterterms were given explicitly and compared to the results obtained recently in  Refs. \cite{ulf,ulf2}. Applications to the isospin breaking in the light quark condensates and to the pion masses were briefly discussed. Upon using naive dimensional estimates for the electromagnetic counterterm, only small corrections to the situation for $e=0$ were obtained for the latter.

At next-to-leading order, the complete expression of the scattering amplitude with a pair of charged pions has been worked out. The counterterms are required both to absorb the divergences produced by the virtual photon loops, and to provide a gauge invariant result. Actually, the second part of this statement was not checked explicitly. It would require that the gauge dependence of the various low energy constants were analysed in detail. The S-wave scattering lengths for the processes $\pi^0\pi^0\rightarrow\pi^0\pi^0$ and $\pi^+\pi^-\rightarrow\pi^0\pi^0$ 
have been computed. For the latter, a proper definition requires that one 
considers the process with the emission of a single soft photon, in order to obtain an infrared finite cross section. The results for the electromagnetic shifts in the scattering lengths at next-to-leading order,
\bea
\Delta a_0(00;00) &=& (-3.2\pm 0.1)\times 10^{-3}
\nonumber\\
\Delta a_0(+-;00) &=& (-1.2\pm 0.7)\times 10^{-3}\ ,
\nonumber
\ena
confirm the conclusion reached at the end of the leading order analysis~: Although these numbers are small as compared to the corresponding values of the one loop strong scattering lentghs, which are equal to $a_0(00;00)\vert_{\rm str}=0.039$ and $a_0(+-;00)\vert_{\rm str}=-0.082$, respectively, the electromagnetic corrections are however comparable in size to the two loop strong interaction effects. 
 In both cases the contributions from the counterterms $k_i$, estimated as mentioned above, are reflected by the uncertainties in these numbers and thus are seen to remain numerically small. 
Although the analysis at next-to-leading order has been carried out in the standard case, it is our feeling, based on the discussion at the end of Section 2, that the conclusion concerning the size of electromagnetic corrections to the S-wave scattering lengths will also hold within the generalized framework. Hopefully, an explicit calculation will confirm this expectation in the future.

As a final remark, we ought to stress that the procedure followed in order to define the scattering length $a_0(+-;00)$ in Section 5 would be rather natural if we had direct experimental access to $\pi^+-\pi^-$ scattering. This, however, is not the case. Instead, one has to proceed along indirect ways, like for instance $K_{\ell 4}$ decays. This process, involving also leptons, has radiative corrections of its own, and which are only partly covered by the analysis we have presented here. Thus, they require a specific treatment, which should preferably be done in a systematic approach. This in turn means that the framework described in Sections 2 and 3 has to be extended, in order to accomodate the leptons. This next step, however, is beyond the scope of the present paper. Our purpose here was rather to assess the typical size of electromagnetic corrections, given a sensible, but not necessarily unique, definition of scattering lengths, and to exhibit, as illustrated in Figs. 2 and 3 for instance, some of the new features that emerge in low energy $\pi-\pi$ scattering once electromagnetic interactions are taken into account.

\indent

\noindent
{\large{\bf Acknowledgements}}

We thank J.~Gasser and J.~Stern for suggesting to undertake this analysis, and for informative discussions. For further discussions and/or correspondance, we thank B.~Moussallam, H.~Sazdjian and S.~Steininger, as well as M.~Perrottet and E.~de~Rafael for sharing their insights on various aspects of radiative corrections. We are grateful to J.~Kambor and to D.~Wyler for extending to us the warm hospitality of the Institute of Theoretical Physics of the University of Z{\"u}rich, where part of this work was done.

\newpage

\noindent
{\Large{\bf Appendix}}

\renewcommand{\theequation}{A.\arabic{equation}}
\setcounter{equation}{0}

\indent

\noindent
{\it A1. Divergent Part of the One Loop Functional}

\indent

At next-to-leading order, the generating functional is
\eq
e^{i{\cal Z}(v_{\mu},a_{\mu},s,p,Q_L,Q_R)}=N\int\left[dU\right]\left[dA_{\mu}\right]
e^{i\int d^{4}x \left\{{\cal L}^{(2)}+{\cal L}^{(4)}\right\}},
\en
where the integration over the fields is carried out in the one loop
approximation. To evaluate the divergent part of the one loop
functional, we transform the lagrangian ${\cal L}^{(2)}$ of
\rf{p2} to Euclidian spacetime,
${\cal L}^{(2)}\rightarrow{\cal L}^{(2)}_{ E}$, and expand 
the fields $U$ and $A_{\mu}$ around the classical 
solutions $\bar{U}$ and $\bar{A}_{\mu}$ of the equations of motion,
\bea
\lbl{fluq}
U&=&ue^{i\xi/F}u=u\left({\bf 1}+i\frac{\xi}{F}
-\frac{1}{2}\frac{\xi^{2}}{F}+\cdots\right)u
\nonumber\\
&=&\bar{U}+\frac{i}{{F}}u\xi u-\frac{1}{2F^2}u\xi^{2}u+\cdots\nonumber\\[2mm]
A_{\mu}&=&\bar{A}_{\mu}+\epsilon_{\mu},
\ena
where $\bar{U}=u^{2}$, and $\xi$ is a traceless hermitian matrix. 
We next insert the expansion \rf{fluq} into the action ${\cal S}_E$, and keep only the terms which are at most quadratic in the quantum fluctuations of the meson and of the photon fields. Collecting these fluctuations in $\eta^{A}$, where $A$ runs from 1 to $N_f^2+3$, 
$\eta=(\xi^{1},\ldots ,\xi^{N_f^2-1},\epsilon^{0},\ldots,\epsilon^{3})$
we obtain, at the one loop level,
\eq
{\cal S}_E|_{\rm one\;loop}=\frac{1}{2}\int d^{4}x_{E}\;
\bigg\{\, \eta^{A}\left(
-\Sigma_{\mu}\Sigma_{\mu}\delta^{AB}+\Lambda^{AB}\right)\eta^{B}
\,-\,\left({{a-1}\over a}\right)\,
\epsilon_{\mu}\partial_{\mu}\partial_{\nu}\epsilon_{\nu}
\,\bigg\}\ ,\lbl{SE}
\en
where $\Sigma_{\mu}$ and $\Lambda^{AB}$ are $(N_f^2+3)\times(N_f^2+3)$ 
matrices,
\bea
\Sigma_{\mu}&=&\partial_{\mu}{\bf 1}+
\left( \begin{array}{cc}
\Gamma^{ab}_{\mu}&X_{\mu}^{a\rho}\\
&\\
X_{\mu}^{\sigma b}&0
\end{array} \right)
=\partial_{\mu}{\bf 1}+Y_{\mu},\nonumber\\
\nonumber\\[2mm]
\Lambda&=&
\left( \begin{array}{cc}
\sigma^{ab}&\frac{1}{2}\gamma^{a\rho}\\
&\\
\frac{1}{2}\gamma^{\sigma b}&\rho\delta^{\sigma\rho}
\end{array} \right) ,\\
\nonumber
\ena
with
\bea
\sigma^{ab}&=&-\frac{1}{2}\langle[\Delta_{\mu},\lambda^{a}]
[\Delta_{\mu},\lambda^{b}]\rangle
+\frac{1}{4}\langle\sigma\{\lambda^{a},\lambda^{b}\}\rangle
-\frac{1}{4}F^2\langle H_L \lambda^{a}\rangle\langle H_L \lambda^{b}\rangle
\nonumber\\ 
&&-\frac{1}{8}{C\over{F^2}}\langle [H_{R}+H_{L},\lambda^{a}][H_{R}-H_{L}, 
\lambda^{b}]+[H_{R}+H_{L},\lambda^{b}][H_{R}-H_{L},\lambda^{a}]\rangle\ ,
\nonumber\\
\gamma^{a}_{\mu}&=&F\langle\left([H_{R},\Delta_{\mu}]+
\frac{1}{2}D_{\mu}H_{L}\right)\lambda^{a}\rangle\ ,
\nonumber\\
\rho&=&\frac{3}{8}F^2\langle H_{L}^{2}\rangle\ ,
\nonumber\\
\Gamma_{\mu}^{ab}&=&-\frac{1}{2}\langle [\lambda^{a},\lambda^{b}]\Gamma_{\mu}
\rangle\ ,
\nonumber\\
X_{\mu}^{a\rho}&=&-X_{\mu}^{\rho a}=-\frac{1}{4}F\langle H_L \lambda^{a}\rangle
\delta^{\rho}_{\mu}\ ,
\ena
and
\bea
D_{\mu}&=&\partial_{\mu} +[\Gamma_{\mu},\cdot]\ ,
\nonumber\\
\Gamma_{\mu}&=&\frac{1}{2}[u^{+},\partial_{\mu}u]-\frac{1}{2}iu^{+}G_{\mu}^R u-
\frac{1}{2}iuG_{\mu}^L u^{+}\ ,
\nonumber\\
\Delta_{\mu}&=&\frac{1}{2}u^{+}d_{\mu}\bar{U}u^{+}=-\frac{1}{2}ud_{\mu}
\bar{U}^{+}u\ ,
\nonumber\\
\sigma&=&\frac{1}{2}(u^{+}\chi u^{+}+u\chi^{+} u)\ ,
\nonumber\\
G_{\mu}^R&=&v_{\mu} +Q_R {\bar A}_{\mu} +a_{\mu}\ ,
\nonumber\\
G_{\mu}^L&=&v_{\mu} +Q_L {\bar A}_{\mu} -a_{\mu}\ ,
\nonumber\\
H_{R}&=&u^{+}Q_R u+ uQ_L u^{+}\ ,
\nonumber\\
H_{L}&=&u^{+}Q_R u- uQ_L u^{+}\ .
\ena
The expression for $\sigma^{ab}$ as given in Eq. (21) of \cite{res} was not symmetrized with respect to the indices $a$ and $b$, as was pointed out by the authors of Ref. \cite{ulf}. Using the correctly symmetrized expression given above, one finds that the divergences of the counterterms $K_{15,16,17}$ in Eq. (35) of \cite{res} have to be replaced by
\eq
\Sigma_{15} = 3/2+3Z+14Z^2\, ,\ \Sigma_{16} = -3-3Z/2-Z^2\, ,
\ \Sigma_{17}=3/2-3Z/2+5Z^2\ .
\en

In the Feynman gauge $a=1$, we thus obtain a gaussian integral for the
generating functional, so that, up to an irrelevant constant contribution,
\begin{equation}
{\cal Z}_{E}|_{\rm one\;loop}=\frac{1}{2}\ln (\det {\cal D})\ ,
\end{equation}
with ${\cal D}^{AB}=-\Sigma_{\mu}\Sigma_{\mu}\delta^{AB}+
\Lambda^{AB}$.  
To renormalize the determinant we use dimensional regularization. 
In Minkowski spacetime, the one loop functional in $d=4$ dimensions is
given by
\eq\label{zone}
{\cal Z}_{\rm one\;loop}=-\frac{1}{16\pi^{2}}\frac{1}{d-4}\int d^{4}x\; \mbox{Tr}\left(
\frac{1}{12}Y_{\mu\nu}Y^{\mu\nu}+\frac{1}{2}\Lambda^{2}\right) + \mbox{ finite
parts},\lbl{divtr}
\en
where Tr means the trace in the ``flavour'' space $\eta^{A}$ and $Y_{\mu\nu}$
denotes the field strength tensor of $Y_{\mu}$,
\eq
Y_{\mu\nu}=\partial_{\mu}Y_{\nu}-\partial_{\nu}Y_{\mu} +[Y_{\mu},Y_{\nu}].
\en

For $a\neq 1$, we obtain, in Eq. \rf{SE}, a second order differential operator of the so called non minimal type. Heat kernel techniques, based on pseudodifferential operator methods \cite{wid}, have been developped for operators of that type  in the literature (see, for instance, \cite{heat} and references therein). They would allow, for instance, to study the gauge dependence of ${\cal Z}_{\rm one\;loop}$, but we shall not further discuss this point in the present 
article.

\indent

In the Feynman gauge, the divergent part for an arbitrary number of flavours $N_f$ results from a straightforward evaluation of the trace in Eq. \rf{divtr}. If we retrict ourselves to sources satisfying the condition $\langle\,Q_L\,\rangle = \langle\,Q_R\,\rangle =
\langle\,Q\,\rangle$, where Q is the constant charge matrix for $N_f$ flavours, the result becomes\footnote{We have written $U$ and $A$ instead of ${\bar U}$ and ${\bar A}$.}
\bea
&&{1\over 12}\, {\rm Tr}\,\big(\,Y^{\mu\nu}\,Y_{\mu\nu}\,\big)\ +
\ {1\over 2}\, {\rm Tr}\,\big(\,\sigma^2\,\big)\ =
\nonumber\\
&&\qquad\ 
{{N_f}\over 48}\,\langle\,d^{\mu} U d^{\nu} U^+ d_{\mu} U d_{\nu}U^+\,\rangle
+{{N_f}\over 24}\,\langle\,d^{\mu} U^+ d_{\mu} U d^{\nu} U^+ d_{\nu}U\,\rangle
\nonumber\\
&&\qquad
+{1\over 8}\,\langle\,d^{\mu} U^+ d^{\nu} U\,\rangle
\,\langle d_{\mu} U^+ d_{\nu}U\,\rangle
+{1\over 16}\,\langle\,d^{\mu} U^+ d_{\mu} U\,\rangle
\,\langle d^{\nu} U^+ d_{\nu}U\,\rangle
\nonumber\\
&&\qquad
-i{{N_f}\over 12}\,\langle\,G^R_{\mu\nu} d^{\mu} U d^{\nu} U^+ +
G^L_{\mu\nu} d^{\mu} U^+ d^{\nu} U\,\rangle
\nonumber\\
&&\qquad
- {{N_f}\over 12}\,\langle\,G^R_{\mu\nu} U G^{L\,\mu\nu} U^+\,\rangle
- {{N_f}\over 24}\,\langle\, G^R_{\mu\nu} G^{R\,\mu\nu} +
G^L_{\mu\nu} G^{L\,\mu\nu}\,\rangle
\nonumber\\
&&\qquad
+{{N_f}\over 8}\,\langle\,d^{\mu} U^+ d_{\mu} U(\chi^+U+U^+\chi)\,\rangle
+{1\over 8}\,\langle\,d^{\mu} U^+ d_{\mu} U\,\rangle
\langle\,\chi^+U+U^+\chi\,\rangle
\nonumber\\
&&\qquad
+{{N_f^2-4}\over{16N_f}}\,\langle\,\chi^+U  \chi^+U+U^+\chi U^+\chi\,\rangle
\nonumber\\
&&\qquad
+{{N_f^2+2}\over{16N_f^2}}\,\langle\,\chi^+U+U^+\chi\,\rangle^2
+{{N_f^2-4}\over{8N_f}}\,\langle\,\chi^+\chi\,\rangle
\nonumber\\
&&\qquad
+{1\over 6}\,\langle\,Q\,\rangle^2\,F^{\mu\nu}F_{\mu\nu}
\nonumber\\
&&\qquad
-{{3F^2}\over 4}\,\langle\,d^{\mu}U^+Q_RUd_{\mu}U^+Q_RU +
d^{\mu}UQ_LU^+d_{\mu}UQ_LU^+\,\rangle
\nonumber\\
&&\qquad
-{{3F^2}\over 4}\,\langle\,d^{\mu}Ud_{\mu}U^+Q_R^2 +
d^{\mu}U^+d_{\mu}UQ_L^2\,\rangle
\nonumber\\
&&\qquad
-\bigg( {1\over 4} - {{N_fZ}\over 2}\bigg)F^2\,\langle\,
d^{\mu}U^+d_{\mu}UQ_LU^+Q_RU + d^{\mu}Ud_{\mu}U^+Q_RUQ_LU^+\,\rangle
\nonumber\\
&&\qquad
+ZF^2\,\langle\,d^{\mu}U^+d_{\mu}U\,\rangle
\langle\,Q_RUQ_LU^+\,\rangle
-ZF^2\,\langle\,d^{\mu}U^+d_{\mu}UQ_LU^+ + d^{\mu}Ud_{\mu}U^+Q_RU\,\rangle
\,\langle\,Q\,\rangle\,\nonumber\\
&&\qquad
+{{F^2}\over 2}\,\langle\,
d^{\mu}U^+Q_Rd_{\mu}UQ_L\,\rangle
+2ZF^2\,\langle\,d^{\mu}U^+Q_RU\,\rangle \,
\langle\,d_{\mu}UQ_LU^+\,\rangle
\nonumber\\
&&\qquad
-{{F^2}\over 4}\,\langle\,(U\chi^+ + \chi U^+)Q_R^2
+(U^+\chi +\chi^+U)Q_L^2\,\rangle
\nonumber\\
&&\qquad
+\bigg({1\over 4} + {{N_fZ}\over 2}\bigg)F^2\,
\langle\,\chi^+Q_RUQ_L + \chi Q_L U^+ Q_R\,\rangle
\nonumber\\
&&\qquad
+\bigg({1\over 4} + {{N_fZ}\over 2}\bigg)F^2\,
\langle\,\chi^+UQ_LU^+Q_RU + \chi U^+Q_RUQ_LU^+\,\rangle
\nonumber\\
&&\qquad
-ZF^2\,\langle\,(U\chi^+ + \chi U^+)Q_R + (\chi^+U + U^+\chi )Q_L\,\rangle
\,\langle\,Q\,\rangle
\nonumber\\
&&\qquad
+ZF^2\,\langle\,\chi^+U + U^+\chi\,\rangle\,
\langle\,Q_RUQ_LU^+\,\rangle
\nonumber\\
&&\qquad
+{{F^2}\over 4}\,\langle\,d_{\mu}U^+[(c_R^{\mu}Q_R),Q_R]U
+d_{\mu}U[(c_L^{\mu}Q_L),Q_L]U^+\,\rangle
\nonumber\\
&&\qquad
+{{F^2}\over 4}\,\langle\,d_{\mu}U^+Q_RU(c_L^{\mu}Q_L)
+d_{\mu}UQ_LU^+(c_R^{\mu}Q_R)\,\rangle
\nonumber\\
&&\qquad
+\bigg(2Z + 2N_fZ^2\bigg)F^4\,\langle\,(Q_RUQ_LU^+)^2\,\rangle
-\bigg(2Z - 2N_fZ^2\bigg)\,\langle\,Q_R^2UQ_L^2U^+\,\rangle
\nonumber\\
&&\qquad
-8Z^2F^4\,\langle\,Q_L^2U^+Q_RU + Q_R^2UQ_LU^+\,\rangle\,
\langle\,Q\,\rangle
\nonumber\\
&&\qquad
+\bigg({3\over 2} + 8Z^2\bigg)F^4\,\langle\,Q_RUQ_LU^+\,
\rangle^2
-{3{F^4}\over 2}\,\langle\,Q_R^2+Q_L^2\,\rangle\,
\langle\,Q_RUQ_LU^+\,\rangle
\nonumber\\
&&\qquad
+\bigg({3\over 8} + Z^2\bigg)F^4\,
\langle\,Q_R^2+Q_L^2\,\rangle^2
-Z^2F^4\langle\,Q_R^2-Q_L^2\,\rangle^2\ ,\lbl{divNf}
\ena
where the covariant derivatives of the sources $Q_L(x)$ and $Q_R(x)$ are
\eq
c^{I}_{\mu}Q_{I}=\partial_{\mu} Q_{I}-i[G^{I}_{\mu},Q_{I}]\ ,
\hspace{2cm}I=R,L\ ,
\en
whereas $G^R_{\mu\nu}$ and $G^L_{\mu\nu}$ are the field strength tensors of $G^R_{\mu}$ and $G^L_{\mu}$,
respectively, 
\eq\label{gmu}
G^I_{\mu\nu} \ =\ {\partial}_{\mu} G^I_{\nu} - {\partial}_{\nu} G^I_{\mu} 
- i \left[ G^I_{\mu} ,
G^I_{\nu} \right]\ , \hspace{2cm}I=R,L\ .
\en

\indent

For $N_f=2$, and after having used the equations of motion and the
trace identities for $2\times 2$ matrices, the above expression
reduces to
\bea
&&{1\over 12}\, {\rm Tr}\,\big(\,Y^{\mu\nu}\,Y_{\mu\nu}\,\big)\ +
\ {1\over 2}\, {\rm Tr}\,\big(\,\sigma^2\,\big)\ =
\nonumber\\
&&\ \ \qquad
{1\over 12}\,\langle\,d^{\mu} U^+ d_{\mu} U\,\rangle^2
+{1\over 6}\,\langle d^{\mu} U^+ d^{\nu}U\,\rangle\,
\langle\,d_{\mu} U^+ d_{\nu} U\,\rangle
\nonumber\\
&&\qquad
-{1\over{32}}\,\langle\,\chi^+U+U^+\chi\,\rangle^2
+{1\over 2}\,\langle\,d^{\mu}U^+d_{\mu}\chi 
+ d^{\mu}\chi^+d_{\mu}U\,\rangle
\nonumber\\
&&\qquad
-{1\over 6}\,\langle\,G^R_{\mu\nu} U G^{L\,\mu\nu}U^+\,\rangle
-{i\over 6}\,\langle\,G^R_{\mu\nu} d^{\mu} U d^{\nu} U^+ +
G^L_{\mu\nu} d^{\mu} U^+ d^{\nu} U\,\rangle
\nonumber\\
&&\qquad
+{1\over 2}\,\langle\,\chi^+\chi\,\rangle
-{1\over 12}\,\langle\, G^R_{\mu\nu} G^{R\,\mu\nu} +
G^L_{\mu\nu} G^{L\,\mu\nu}\,\rangle
+Re ({\rm det}\chi)
\nonumber\\
&&\qquad
+{1\over 6}\,\langle\,Q\,\rangle^2\,F^{\mu\nu}F_{\mu\nu}
\nonumber\\
&&\qquad
-{{3F^2}\over 4}\,\langle\,d_{\mu} U^+ d_{\mu} U\,\rangle
\,\langle\,Q_R^2 + Q_L^2\,\rangle
+\bigg({3\over 4}-Z\bigg)F^2\,\langle\,d^{\mu} U^+ d_{\mu} U\,\rangle
\,\langle\,Q\,\rangle^2
\nonumber\\
&&\qquad
+2ZF^2\,\langle\,d^{\mu}U^+d_{\mu}U\,\rangle
\langle\,Q_RUQ_LU^+\,\rangle
\nonumber\\
&&\qquad
-{{3F^2}\over 4}\,
\big(\,\langle\,d^{\mu}U^+Q_RU\,\rangle\,
\langle\,d_{\mu}U^+Q_RU\,\rangle\,+
\,\langle\,d^{\mu}U^+Q_LU^+\,\rangle\,
\langle\,d_{\mu}UQ_LU^+\,\rangle\,\big)
\nonumber\\
&&\qquad
+2ZF^2\,\langle\,d^{\mu}U^+Q_RU\,\rangle\,
\langle\,d_{\mu}UQ_LU^+\rangle
-{{F^2}\over 8}\,\langle\,\chi^+U+U^+\chi\,\rangle
\,\langle\,Q_R^2+Q_L^2\,\rangle
\nonumber\\
&&\qquad
-ZF^2\,\langle\,\chi^+U+U^+\chi\,\rangle\,\langle\,Q\,\rangle^2
+\bigg({1\over 4}+2Z\bigg)F^2
\,\langle\,\chi^+U+U^+\chi\,\rangle\,
\langle\,Q_RUQ_LU^+\,\rangle
\nonumber\\
&&\qquad
+\bigg({1\over 8}-Z\bigg)F^2
\,\langle\,(\chi U^+ - U\chi^+ )Q_RUQ_LU^+
+(\chi^+U-U^+\chi )Q_LU^+Q_RU\,\rangle
\nonumber\\
&&\qquad
+{{F^2}\over 4}\,\langle\,d_{\mu}U^+[(c_R^{\mu}Q_R),Q_R]U
+d_{\mu}U[(c_L^{\mu}Q_L),Q_L]U^+\,\rangle
\nonumber\\
&&\qquad
+{{F^2}\over 4}\,\langle\,d_{\mu}U^+Q_RU(c_L^{\mu}Q_L)
+d_{\mu}UQ_LU^+(c_R^{\mu}Q_R)\,\rangle
\nonumber\\
&&\qquad
+\bigg({3\over 2}+3Z+12Z^2\bigg)F^4\,
\langle\,Q_RUQ_LU^+\,\rangle^2
-{{3F^4}\over 2}\,\langle\,Q_RUQ_LU^+\,\rangle
\,\langle\,Q_R^2+Q_L^2\,\rangle
\nonumber\\
&&\qquad
-\bigg(3Z+12Z^2\bigg)F^4\,\langle\,Q_RUQ_LU^+\,\rangle\,
\langle\,Q\,\rangle^2
+\bigg({3\over 8}-{3Z\over 4}+Z^2\bigg)F^4\,
\langle\,Q_R^2+Q_L^2\,\rangle^2
\nonumber\\
&&\qquad
+\bigg({{3Z}\over 2}-2Z^2\bigg)F^4\,\langle\,Q_R^2+Q_L^2\,\rangle
\,\langle\,Q\,\rangle^2
-Z^2F^4\langle\,Q_R^2-Q_L^2\,\rangle^2
+4Z^2F^4\,\langle\,Q\,\rangle^4\, .
\ena

\indent

\noindent
{\it A2. Loop Functions}

\indent

We recall the definition of the function ${\bar J}_{PQ}(s)\equiv J_{PQ}(s)-
J_{PQ}(0)$, with
\eq
J_{PQ}(s)\ =\ -i\,\int\,{{d^dq}\over{(2\pi)^d}}\,
{1\over{(q^2-M_P^2+i\epsilon)[(q-p)^2-M_Q^2+i\epsilon]}}\ ,
\en 
and $s=p^2$.
For $s>(M_P+M_Q)^2$ and for $d=4$, one finds
\bea
32\pi^2\,{\bar J}_{PQ}(s) &=& 2+{{\Delta_{PQ}}\over s}\,
\ln\,{{M_Q^2}\over{M_P^2}}-
{{\Sigma_{PQ}}\over {\Delta_{PQ}}}
\,\ln\,{{M_Q^2}\over{M_P^2}}
\nonumber\\
&&
+{{\lambda_{PQ}^{1\over 2}(s)}\over s}\,
\ln\,\left[\,{{(s-\lambda_{PQ}^{1\over 2}(s))^2-\Delta_{PQ}^2}\over
              {(s+\lambda_{PQ}^{1\over 2}(s))^2-\Delta_{PQ}^2}}\,\right]
+2i\pi{{\lambda_{PQ}^{1\over 2}(s)}\over s}\ ,
\ena
with
\eq
\Sigma_{PQ}\ =\ M_P^2+M_Q^2\ ,\ \ \Delta_{PQ}\ =\ M_P^2-M_Q^2\ ,
\en
and
\eq
\lambda_{PQ}(s)\ =\ \lambda (s,M_P^2,M_Q^2)\ .
\en
For the threshold expansions of the amplitudes or cross sections, the following expressions have been used,
\bea
&&Re\,\Joo(4\Mpic)\ =\ {1\over{8\pi^2}}\,\bigg[\,
1-{{\Delta_{\pi}}\over{\Mpin}}+{2\over3}\,{{\Delta_{\pi}^2}\over{M_{\pi^0}^4}}
+\cdots\,\bigg]\ ,
\nonumber\\
&&Re\,\Jpm(4\Mpin)\ =\ {1\over{8\pi^2}}\,\bigg[\,
1+{{\Delta_{\pi}}\over{\Mpin}}-{1\over3}\,{{\Delta_{\pi}^2}\over{M_{\pi^0}^4}}
+\cdots\,\bigg]\ ,
\nonumber\\
&&Re\,{\bar J}_{+0}(-\Delta_{\pi})\ =\ -\,{1\over{96\pi^2}}\,
{{\Delta_{\pi}}\over{\Mpin}}\,\bigg[\,1-{3\over 5}\,
{{\Delta_{\pi}}\over{\Mpin}}+\cdots\,\bigg]\ ,
\nonumber\\
&&Re\,{\bar J}_{+0}\big( (M_{\pi^{\pm}}+M_{\pi^0})^2\big)\ =
\ {1\over{8\pi^2}}\,\bigg[\,
1-{1\over 48}\,{{\Delta_{\pi}^2}\over{M_{\pi^0}^4}}+\cdots\,\bigg]\ ,
\nonumber\\
&&Re\,{\bar J}_{+0}\big( (M_{\pi^{\pm}}-M_{\pi^0})^2\big)\ =
\ {1\over{384\pi^2}}\,{{\Delta_{\pi}^2}\over{M_{\pi^0}^4}}+\cdots
\ ,
\ena
where only the contributions up to order $O(\Delta_{\pi}^2/M_{\pi^0}^4)$ have been kept.

As far as the function $G_{+-\gamma}(s)$ defined in Eq. \rf{Gpmg} is concerned,
introducing two Feynman parameters and performing standard manipulations leads to the following integral representation
\eq
G_{+-\gamma}(s)\ =\ -\,{1\over{32\pi^2}}\,\int_0^1\,dx\,\int_0^1\,dy\,
{1\over{f(x)}}\cdot{d\over{dy}}\,\ln\bigg[\,
y^2f(x)+(1-y)m_{\gamma}^2\,\bigg]\ +\ O(m_{\gamma}^2)\ ,
\en
where
\eq
f(x)\ =\ \Mpic-x(1-x)s-i\epsilon\ .
\en
For $s<0$, the roots of $f(x)$ are real and lie outside of the interval $[\,0,1\,]$, so that the integration is straightforward,
\bea
G_{+-\gamma}(s) &=& -{1\over{32\pi^2s\sigma}}\,\bigg\{\,
4\,{\rm Li}_2\left({{1-\sigma}\over{1+\sigma}}\right)
+{{\pi^2}\over3}+\ln^2\left({{\sigma-1}\over{\sigma+1}}\right)
\nonumber\\
&&\qquad
+2\,\bigg[\,\ln\left({-s\over{\Mpic}}\right)
-\ln\left({m_{\gamma}^2\over{\Mpic}}\right)
+2\ln(\sigma)\,\bigg]\,
\ln\left({{\sigma-1}\over{\sigma+1}}\right)\,\bigg\}
\ .
\ena
The dilogarithm or Spence function is defined as usual,
\eq
{\rm Li}_2(x)\ =\ -\,\int_1^x\,{{\ln t}\over{1-t}}\,dt\ .
\en
The expression \rf{Gpmg2} for $s>4\Mpic$ follows by analytic continuation with the $i\epsilon$ prescription.

\newpage

\noindent
{\it A3. Radiative $\pi-\pi$ Scattering}

\indent

Let
\eq
f({\cal E},s)\ =\ \sqrt{{{s-2{\cal E}\sqrt{s}-4\Mpin}\over{s-4\Mpin}}}\cdot
\sqrt{{s\over{s-2{\cal E}\sqrt{s}}}}\ .
\en
Then the function $F(s;\Delta E)$, resulting from the integration over the energy of the emitted photon, is given by the indefinite integral
\bea
&&
F(s;{\Delta E})\ =\ \int^{\Delta E}
\,{{d{\cal E}}\over{\cal E}}\,[\,(s-2{\cal E}\sqrt{s}-\Mpin)^2\,
f({\cal E},s)-(s-\Mpin)^2\,]=
\nonumber\\
&&-\ {1\over2}(s-2{\Delta E}\sqrt{s})(6\Mpin+2{\Delta E}\sqrt{s}-3s)
f({\Delta E},s) \nonumber\\
&&\ -(s-\Mpin)^2\,\ln\bigg\{\,
\sqrt{s}\,\big[\,s-2{\Delta E}\sqrt{s}-4\Mpin\,\big]\,
\big[\,1+f({\Delta E},s)\,\big]\,+\,
4\Mpin{\Delta E}\,\big[\,1+2f({\Delta E},s)\,\big]\,\bigg\}
\nonumber\\
&&-\ \sqrt{{s\over{s-4\Mpin}}}\,(s^2-4s\Mpin+3M_{\pi^0}^4)
\\
&&\qquad
\ln\bigg\{\,(s-4\Mpin)\,
\bigg[\,(s-2{\Delta E}\sqrt{s})
\big[\,1-\sqrt{{{s-4\Mpin}\over{s}}}f({\Delta E},s)\,\big]
-2\Mpin\,\bigg]\,
\bigg\}\ +\ {\rm Cst}\ .\nonumber
\ena
For the threshold expansion of the total cross section, we need 
\bea
&&F(4\Mpic ; {{\Delta_{\pi}}\over{M_{\pi^{\pm}}}})
\,-\,F(4\Mpic ;0) \ =\ \nonumber\\
&&\qquad\qquad\qquad
12\Mpic(\Mpin-2\Mpic)+(4\Mpic-\Mpin)^2\,\ln\left({{4\Mpic}\over{\Mpin}}\right)
\nonumber\\
&&\qquad\qquad\qquad
+{{M_{\pi^{\pm}}}\over{\sqrt{\Delta_{\pi}}}}\,(16M_{\pi^{\pm}}^4-
16\Mpic\Mpin+3M_{\pi^0}^4)\,
\ln\left({{\Mpin+2\Delta_{\pi}-2M_{\pi^{\pm}}{\sqrt{\Delta_{\pi}}}}
\over{\Mpin}}\right)
\nonumber\\
&&\qquad\qquad\ =\  18(\ln 2-1)\,M_{\pi^0}^4\,+\cdots
\ena
Since the difference $F(4\Mpic ; {{\Delta_{\pi}}/{M_{\pi^{\pm}}}})
-F(4\Mpic ;0)$ already appears multiplied by $e^2\Delta_{\pi}$ in the expression \rf{sgamma} of ${\tilde\sigma}^{+-;00\gamma}$ at threshold, we have dropped, in the last line of (A.26), all the contributions that vanish for ${\Delta_{\pi}}\to 0$.

\end{document}